\documentclass[graybox]{svmult}

\usepackage[toc]{appendix}

\usepackage{type1cm}        
\usepackage{makeidx}         
\usepackage{graphicx}        
\usepackage{multicol}        
\usepackage[bottom]{footmisc}

\usepackage{newtxtext}       %
\usepackage{newtxmath}       
\usepackage{combelow}        
\usepackage{url}
\usepackage{bm}
\usepackage{braket}          
\usepackage{slashed}         
\usepackage{nicefrac}
\makeindex             
                       
\newcommand{\onehalf}{{\nicefrac{1}{2}}}

\begin{document}

\title*{Particle polarization, spin tensor and the Wigner distribution in relativistic systems}
\titlerunning{Relation between particle polarization and the spin tensor}
\author{Leonardo Tinti \and Wojciech Florkowski}
\authorrunning{Leonardo Tinti \and Woiciech Florkowski}
\institute{Leonardo Tinti \at Institut f\"ur Theoretische Physik, Johann Wolfgang Goethe-Universit\"at, Max-von-Laue-Str.~1, D-60438 Frankfurt am Main, Germany \\ \email{dr.leonardo.tinti@gmail.com}
\and Wojciech Florkowski \at  
Jagiellonian University, ul. prof. St. Łojasiewicza 11, 30-348 Kraków, Poland \\ \email{wojciech.florkowski@uj.edu.pl}}

\maketitle

\abstract{Particle spin polarization is known to be linked both to rotation (angular momentum) and magnetization of a many particle system. However, in the most common formulation of relativistic kinetic theory, the spin degrees of freedom appear only as degeneracy factors multiplying phase-space distributions. Thus, it is important to develop theoretical tools that allow to make predictions regarding the spin polarization of particles, which can be directly confronted with experimental data. Herein, we discuss a link between the relativistic spin tensor and particle spin polarization, and elucidate the connections between the Wigner function and average polarization. Our results may be useful for theoretical interpretation of heavy-ion data on spin polarization of the produced hadrons.}

\section{Introduction}
\label{sec:intro}

In relativistic heavy-ion collisions the produced matter is formed at extreme conditions of high temperature and density~\cite{Busza:2018rrf}. The exact details of the evolution of  the resulting fireball are difficult to track, however, some generally accepted concepts are typically assumed now. In particular, the evidence has been found that the strongly interacting matter behaves as an almost perfect (low viscosity) fluid (for recent reviews see, for example, Refs.~\cite{Romatschke:2017ejr,Florkowski:2017olj}). During its expansion, when the system is diluted enough, the matter undergoes a cross-over phase transition from a strongly interacting quark-gluon plasma to a hadron gas (for systems with negligible baryon number density~\cite{Borsanyi:2010bp,Bazavov:2018mes}). Shortly thereafter, the system becomes too dilute to be properly treated as a fluid, the interaction effectively ends, and produced particles freely stream to the detectors (freeze-out).

The main purpose of relativistic hydrodynamics is to solve the four-momentum conservation equation, $\partial_\mu T^{\mu\nu}=0$, and the (baryon) charge conservation equation, $\partial_\mu J^\mu = 0$, under some realistic approximations (i.e., with suitably chosen initial conditions given by the Glauber model~\cite{Miller:2007ri} or the theory of color glass condensate~\cite{Gelis:2010nm}, and with a realistic equation of state obtained from an interpolation between the lattice QCD simulations and hadron resonance gas calculations).

In this way one obtains the four-momentum and charge fluxes at freeze-out. These are space-time densities, which are not directly connected with the momentum distributions (and polarization) measured by different experiments. The most common way to link the stress energy tensor and the charge flux at the freeze-out to particle spectra makes use of a classical intuition~\cite{Cooper:1974mv}; namely, one assumes that after the freeze-out the system is described well enough by the distribution functions $f(x,{\bf p})$ for (non interacting) particles and the corresponding functions $\bar f (x,{\bf p})$ for antiparticles. Matching the stress energy tensor and charge current with the corresponding formulas from the relativistic kinetic theory, one can guess the form of the distribution functions, and from that predict the final spectra.

The purpose of this contribution is to extend this formalism to include particle spin polarization (for particles with spin 1/2). In the next section we show the limitations of the traditional kinetic theory in relation to particle polarization degrees of freedom. In Sec. 3 we introduce the concept of the relativistic spin tensor and discuss its relation to particle polarization for a free Fermi-field. In Sec.~4 we show that the appropriate generalization of the distribution function is the Wigner distribution. We summarize and conclude in Sec.~5. Some useful expressions and transformations are given in Appendix~A. For complementary information we refer to the recent reviews~\cite{Wang:2017jpl,Huang:2020xyr,Becattini:2020ngo,Becattini:2020sww}.

\section{Relativistic kinetic theory and its limitations}
\label{sec:RKT}

In the classical relativistic kinetic theory the charge current density $J^\mu$ and the  stress energy tensor $T^{\mu\nu}$ have a rather simple connection with the phase-space distributions of particles and antiparticles~\cite{de_Groot,Florkowski:2010zz},
\begin{equation}\label{kinetic}
 \begin{split}
  J^{\mu} &= \frac{g_S}{(2\pi)^3}\int \frac{d^3 p}{E_{\bf p}} \; p^\mu \, \left( \vphantom{\frac{}{}} f(x,{\bf p}) - \bar f(x,{\bf p})\right), \\
  T^{\mu\nu} &= \frac{g_S}{(2\pi)^3}\int \frac{d^3 p} {E_{\bf p}} \; p^\mu p^\nu \left( \vphantom{\frac{}{}} f(x,{\bf p}) + \bar f(x,{\bf p})\right).
 \end{split}
\end{equation}
Here $g_S=2S+1$ is the degeneracy factor with $S$ being the spin of (massive) particles. In order to properly take into account the polarizaton degrees of freedom, one can easily notice that the framework based on Eqs.~(\ref{kinetic}) should be extended to a matrix formalism. This is so, since the standard kinetic theory essentially assumes equipartition of various spin states. 

In general, the expectation values of the charge current and the stress energy tensor, determined in a generic state of the system described by the density matrix $\rho$, cannot be expressed by the integrals of the form~(\ref{kinetic}) with the integrands depending on a single momentum variable. A generic density matrix can be written as
\begin{equation}\label{density_matrix}
 \rho = \sum_i {\sf P}_i \left| \psi_i \right\rangle \left\langle  \psi_i\right|,
\end{equation}
where ${\sf P}_i$ are classical (non-interfering) probabilities normalized to one, $\sum_i{\sf P}_i=1$, and $|\psi_i\rangle$ are generic quantum states. We assume that $\left\langle \psi_i \right| \left. \!\psi_i\right\rangle=1$ and stress that  $ \left|\,\, \!\psi_i\right\rangle$'s are not necessarily eigenstates of the total energy, linear momentum, angular momentum, or charge operators.

Starting from the definition of the charge current operator
\begin{equation}
 \hat J^\mu (x) = \bar \Psi(x)\gamma^\mu \Psi(x),
\end{equation}
expressed by the non-interacting Fermi fields $\Psi$,
\begin{equation}
 \Psi(x) = \sum_r \int\frac{d^3 p}{(2\pi)^3 \sqrt{2E_{\bf p}}}\left[ \vphantom{\frac{}{}} U_r({\bf p}) a_r({\bf p}) e^{-i p\cdot x} + V_r({\bf p}) b^\dagger_r({\bf p}) e^{i p\cdot x} \right],
\end{equation}
we obtain the normal-ordered expectation value 
\begin{equation}
 \begin{split}
  J^\mu(x) &= \langle: \hat J^\mu (x):\rangle={\rm tr}\left( \rho : \hat J^\mu (x): \right) =\\
  &= \sum_{r,s}\int \frac{d^3p d^3p^\prime}{(2\pi)^6\sqrt{2E_{\bf p}2E_{{\bf p}^\prime}}}\left[ \vphantom{\frac{}{}} \langle a^\dagger_r({\bf p})a_s({\bf p}^\prime)\rangle \bar U_r({\bf p}) \gamma^\mu U_s({\bf p}^\prime) e^{i(p-p^\prime)\cdot x} \right.  \\
 &\left. \frac{}{}\qquad \qquad\qquad \qquad   \quad - \langle b^\dagger_r({\bf p}) b_s({\bf p}^\prime)\rangle \bar V_s({\bf p}^\prime) \gamma^\mu V_r({\bf p}) e^{i(p-p^\prime)\cdot x}  \right.\\
&\left. \frac{}{}\qquad \qquad\qquad \qquad   \quad + \langle a^\dagger_r({\bf p}) b^\dagger_s({\bf p}^\prime)\rangle \bar U_r({\bf p}) \gamma^\mu V_s({\bf p}^\prime) e^{i(p+p^\prime)\cdot x}  \right. \\
&\left. \frac{}{}\qquad \qquad\qquad \qquad  \quad + \langle b_r({\bf p}) a_s({\bf p}^\prime)\rangle \bar V_r({\bf p}) \gamma^\mu U_s({\bf p}^\prime) e^{-i(p+p^\prime)\cdot x}  \right]. \\
 \end{split}
\end{equation}
It is easy to check that the stress energy tensor has an analogous structure. In general, none of the expectation values of the creation-destruction operators vanishes and the integrals over three-momenta cannot be reduced to the Dirac delta functions.

We use the Wigner representation of the Clifford algebra and we adopt the convention for the massive eigenspinors\footnote{Note that in the Weyl representation of the Clifford algebra a different explicit formula for the massive eigenspinors is typically used, but final results remain the same.}

\begin{equation}\label{U_V}
 \begin{split}
   U_r({\bf p}) &= \sqrt{E_{\bf p}+m}\left(\begin{matrix}
															 & \phi_r \\
															 \frac{{\boldsymbol \sigma} \cdot {\bf p}}{E_{\bf p}+m}&\phi_r
				   										   \end{matrix}\right),\\
   V_r({\bf p}) &= \sqrt{E_{\bf p}+m} \left(\begin{matrix}
															\frac{{\boldsymbol \sigma} \cdot {\bf p}}{E_{\bf p}+m}&\chi_r \\
															&\chi_r
														  \end{matrix}\right),
 \end{split}
\end{equation}
with the two component vectors $\phi_r$ and $\chi_r$ being the eigenstates of the matrix $\sigma_z={\rm diag}(1,-1)$,

\begin{equation}
 \begin{split}
  \phi_1 &= \left(\begin{matrix} 1 \\0 \end{matrix}\right), \qquad  \phi_1 = \left(\begin{matrix} 0 \\1 \end{matrix}\right), \\
  \chi_1 &= \left(\begin{matrix} 0 \\1 \end{matrix}\right), \quad  \chi_2 = -\left(\begin{matrix} 1 \\0 \end{matrix}\right),
 \end{split}
\end{equation}
therefore the normalization of the states $|p,r\rangle$ and the anticommutation relations between the creation-destruction operators read

\begin{equation}
 \begin{split}
 &\{a_s({\bf q}),a^\dagger_r({\bf p})\} =\{b_s({\bf q}),b^\dagger_r({\bf p})\}= (2\pi)^3 \delta_{rs}\delta^3({\bf p}-{\bf q}),\\
 & \{a_s({\bf q}),b_r({\bf p})\}=\{a_s({\bf q}),b_r({\bf p})\}=\{a_s({\bf q}),b_r({\bf p})\}=\{a_s({\bf q}),b_r({\bf p})\}=0, \\
 & |p,r\rangle = \sqrt{2E_{\bf p}} a^\dagger_r({\bf p})|0\rangle \quad\Rightarrow \quad \langle q,s|p,r\rangle = (2\pi)^3\delta_{r,s}\delta^3({\bf p}-{\bf q}).
 \end{split}
\end{equation}

Differently from the current density, the total charge has a similar structure to the one  used in the kinetic theory, 
\begin{equation}\label{charge}
 \int \!\!\! d^3x \; J^0(x) = \int \!\! \frac{d^3 p}{(2\pi)^3} \left[ \sum_r \langle a^\dagger_r({\bf p})a_r({\bf p})\rangle -\sum_r \langle b^\dagger_r({\bf p})b_r({\bf p})\rangle\right]
\end{equation}
which directly comes from the  normalization of the bispinors~(\ref{U_V})
\begin{equation}
 \begin{split}
  U^\dagger_r({\bf p})U_s({\bf p}) &= V^\dagger_r({\bf p})V_s({\bf p}) = 2E_{\bf p}\delta_{rs}, \\
  U^\dagger_r({\bf p})V_s(-{\bf p}) &=0, 
 \end{split}
\end{equation}
and the integral representation of the Dirac delta functions, $\delta^{(3)}({\bf p}\pm {\bf p}^\prime)$, used to perform the volume integrals. Following the same steps, one can compute the total four-momentum
\begin{equation}\label{four_mom}
 \int \!\!\! d^3x \; T^{0\mu}(x) = \int \!\! \frac{d^3 p}{(2\pi)^3} p^\mu\left[ \sum_r \langle a^\dagger_r({\bf p})a_r({\bf p})\rangle + \sum_r \langle b^\dagger_r({\bf p})b_r({\bf p})\rangle\right].
\end{equation}
Equations~(\ref{charge}) and~(\ref{four_mom}) should be compared to the analogous formulas obtained from the kinetic-theory definitions~(\ref{kinetic}),
\begin{equation}
 \begin{split}
  \int \!\!\! d^3x \; J^0 &= \int \frac{d^3 p}{(2\pi)^3} \, \left[ \vphantom{\frac{}{}} g_s\int \!\!\! d^3x f(x,{\bf p}) - g_s \int \!\!\! d^3x \bar f(x,{\bf p})\right], \\
  \int \!\!\! d^3x \; T^{0\mu} &= \int \frac{d^3 p} {(2\pi)^3} \; p^\mu  \left[ \vphantom{\frac{}{}} g_s\int \!\!\! d^3x f(x,{\bf p}) + g_s\int \!\!\! d^3x\bar f(x,{\bf p})\right].
 \end{split}
\end{equation}

It is important to note that for non-interacting, spin $\onehalf$ particles the volume integrals are time independent.  The collisionless Boltzmann equation for particles reads
\begin{equation}
 \partial_\mu \left( p^\mu f(x,{\bf p})\right)=0,
 \label{div}
\end{equation}
hence, the expression $p^\mu f$ is a conserved vector current (of course, the same property holds also for the non-interacting antiparticles described by the function $\bar f(x,{\bf p})$). The integral of the divergence (\ref{div}) over a space time-region vanishes. Therefore, as long as the spatial boundary for the volume integral lies outside of the region in which the distribution function does not vanish, the integral $E_{\bf p}\int\!d^3x f$ remains the same at all times. Massive particles always have a positive $E_{\bf p}$, therefore, the volume integral itself is also  time-independent\footnote{Massless particles have a positive energy for any non vanishing momentum and the situation for them is quite similar.}.

Moreover, the space integral equals the integral over the freeze-out hyper-surface~\cite{Cooper:1974mv}, with $d\Sigma_\mu$ being the generic hyper-surface element (note that $d\Sigma_\mu = (d^3x,0,0,0)$ for the volume integrals in the lab frame)
\begin{equation}
 g_s\int \!\!\!d\Sigma_\mu \, p^\mu f(x,{\bf p})=g_s E_{\bf p} \int \!\!\! d^3x f(x,{\bf p}) = (2\pi)^3 E_{\bf p} \frac{d N}{d^3p}.
\end{equation}
The last term here is the invariant number of particles per momentum cell, which is consistent with the formula for the total number of particles
\begin{equation}
 N = \int\!\!\!d^3 x \int \!\!\frac{d^3 p}{(2\pi)^3} g_s f(x,{\bf p}) = \int\!\!\!d^3 x \int \!\!\frac{d^3 p}{(2\pi)^3E_{\bf p}} \,  E_{\bf p} g_s f(x,{\bf p}).
\end{equation}
We note that the factor $(2\pi)^3$ (to be replaced by $(2\pi \hbar)^3$ if the natural units are not used) is included in the momentum integration measure rather than in the definition of the phase-space distributions, and $\int \!\! d^3 p/E_{\bf p}$ is the Lorentz-covariant momentum integral.

In the general case, it can be proved that the expectation value of the number operator is a non-negative quantity and the sum over the spin states is proportional to the (anti)particle number density in momentum space, hence, 
\begin{equation}
N =  \int\!\!\frac{d^3p}{(2\pi)^3} \sum_r \langle a^\dagger_r({\bf p})a_r({\bf p})\rangle
\end{equation}
for particles, and 
\begin{equation}
{\bar N} =  \int\!\!\frac{d^3p}{(2\pi)^3} \sum_r \langle b^\dagger_r({\bf p})b_r({\bf p})\rangle
\end{equation}
for antiparticles. For more information see Appendix~\ref{app:density}.

Consequently, even if the expectation values of the charge current and the stress-energy tensor cannot be written as momentum integrals of the phase-space distributions, it is possible to have consistent distribution functions for the particles and antiparticles in the sense that they can reproduce the correct invariant momentum (anti)particle densities,
\begin{equation}
 \begin{split}
  g_s\int\!\!\!d^3 x f(x,{\bf p}) &=  \sum_r \langle a^\dagger_r({\bf p})a_r({\bf p})\rangle, \\
  g_s\int\!\!\!d^3 x \bar f(x,{\bf p}) &= \sum_r \langle b^\dagger_r({\bf p})b_r({\bf p})\rangle.
 \end{split}
\end{equation}
For heavy-ion collisions, this implies that for any given $J^\mu$ and $T^{\mu\nu}$ at freeze-out, one can construct a pair of the distribution functions ($f$, $\bar f$)  that provide the same total current, energy, and linear momentum,
\begin{equation}
 \begin{split}
 & \int \!\!\! \frac{d^3 p}{(2\pi)^3} \, \left[ \vphantom{\frac{}{}} g_s\int \!\!\! d^3x f(x,{\bf p}) - g_s \int \!\!\! d^3x \bar f(x,{\bf p})\right] =  \\
 & \qquad \qquad = \int \!\!\! \frac{d^3 p}{(2\pi)^3} \left[ \sum_r \langle a^\dagger_r({\bf p})a_r({\bf p})\rangle -\sum_r \langle b^\dagger_r({\bf p})b_r({\bf p})\rangle\right],
 \end{split}
 \label{eq:cond1}
\end{equation}
\begin{equation}
 \begin{split}
  &\int \!\!\! \frac{d^3 p} {(2\pi)^3} p^\mu  \left[ \vphantom{\frac{}{}} g_s\int \!\!\! d^3x f(x,{\bf p}) + g_s\int \!\!\! d^3x\bar f(x,{\bf p})\right] = \\
  & \qquad \qquad= \int \!\!\! \frac{d^3 p}{(2\pi)^3} p^\mu \left[ \sum_r \langle a^\dagger_r({\bf p})a_r({\bf p})\rangle +\sum_r \langle b^\dagger_r({\bf p})b_r({\bf p})\rangle\right].
 \end{split}
 \label{eq:cond2}
\end{equation}
We note that the distributions functions obtained from the conditions (\ref{eq:cond1}) and (\ref{eq:cond2}) provide the correct total charge and four-momentum if used in Eqs.~(\ref{kinetic}) to define the current density and the energy-momentum tensor. However, 
very different distribution functions may provide (after integration) the same macroscopic quantities. In certain cases, to remove such an ambiguity, one can use additional physical insights. For example, the specific forms of the distribution functions can be introduced for systems being in (local) thermodynamic equilibrium or  close to such a state.

In any case, it is important to note that the total charge is sensitive to the imbalance between particles and antiparticles, and the total  four momentum is sensitive to the momentum distribution of both particles and antiparticles. Therefore, the conditions (\ref{eq:cond1}) and (\ref{eq:cond2}) provide an important constraint on the distribution functions. However, the right-hand sides in (\ref{eq:cond1}) and (\ref{eq:cond2}) are sums over the spin states.  Being insensitive to polarization, they are not useful to check if a given extension of kinetic theory reproduces the average polarization in a satisfactory manner. 

In the next section, we will argue that the relativistic spin tensor is sensitive to particle polarization in a very similar way as the charge current is sensitive to particle-antiparticle imbalance and the stress-energy tensor controls the average particle momentum.

\section{The relativistic spin tensor as a polarization sensitive macroscopic object}

In this section we introduce and discuss in more detail one of the main objects of our interest, namely, the relativistic spin tensor. Although, it is less well known compared to the tensors analyzed in the previous section, we are going to demonstrate that its intuitive understanding as a quantity related to particle's polarization is indeed correct.

The Noether theorem links the symmetries of the action to conserved charges and, to a lesser extent, conserved currents. If the action ${\cal A}$ contains only first order derivatives of the fields $\phi^a(x)$, we can write~\cite{Itzykson:1980rh}

\begin{equation}
 {\cal A}[\phi^a] =\int d^4x {\cal L}(\phi, \partial_\mu \phi, x),
\end{equation}
where ${\cal L}$ is the Lagrangian density. If the action is invariant with respect to an infinitesimal transformation
\begin{equation}
 \begin{split}
  & x^\mu \to \xi^\mu= x^\mu + \epsilon \delta x^\mu, \\
  & \phi^a(x) \to \alpha^a (\xi) = \phi^a(x) +\epsilon\delta\phi^a(x) +\epsilon\delta x^\mu\partial_\mu \phi^a(x),
 \end{split}
\end{equation}
one can extract a conserved current
\begin{eqnarray}
 {\cal Q}^\mu &=& \left[ \frac{\partial {\cal L}}{\partial(\partial_\mu\phi^a)} \partial_\nu\phi^a -{\cal L}\delta^\mu_\nu\right]\delta x^\nu - \frac{\partial {\cal L}}{\partial(\partial_\mu \phi^a)}\left(\delta\phi^a(x) +\delta x^\nu\partial_\nu \phi^a(x) \vphantom{\frac{}{}}\right), \nonumber \\ \nonumber \\
\partial_\mu {\cal Q}^\mu &=& 0,
\label{eq:dQ}
\end{eqnarray}
where the summation over repeated indices is understood. Because of the vanishing divergence, the integral over a space-time region of~(\ref{eq:dQ}) vanishes. Hence, if the field flux at the space boundary vanishes\footnote{The space-time region might be finite, with the fields going to zero at the boundary or infinite, as long as the fields decay fast enough to have a vanishing flux at infinity.}, the space integral of ${\cal Q}^0$ is a constant of motion. For instance, considering the action of a free, massive, spin $\onehalf$ spinor field $\Psi$, 
\begin{equation}
 {\cal A}=\int d^4 x \left[ \frac{i}{2}\bar\Psi(x) \gamma^\mu \stackrel{\leftrightarrow}{\partial}_\mu \Psi(x) - m \bar \Psi(x)\Psi(x) \right],
\end{equation}
one has the internal symmetry under a global phase change of the fields with: $\delta x^\mu\equiv 0$, $\delta \Psi = i\epsilon \Psi$, and $\delta \bar\Psi = -i\epsilon \bar \Psi$. The corresponding current in this case is the charge current $J^\mu$,
\begin{equation}
  J^\mu (x) = \bar \Psi(x)\gamma^\mu \Psi(x).
\end{equation}
The invariance under space-time translations (with $\delta x^\mu$ being a constant and $\alpha(\xi)-\phi(x) = -\delta x^\mu\partial_\mu \phi$) yields the canonical stress-energy tensor $T^{\mu\nu}_c$ as the conserved current. The conserved charge in this case is the total four-momentum of the system
\begin{equation}\label{T_c_Dirac}
    T_c^{\mu\nu}(x) = \frac{i}{2}\bar \Psi(x)\gamma^\mu\stackrel{\leftrightarrow}{\partial^\nu} \Psi(x) -g^{\mu\nu} {\cal L} \equiv  \frac{i}{2}\bar \Psi(x)\gamma^\mu\stackrel{\leftrightarrow}{\partial^\nu} \Psi(x).
\end{equation}
In the last passage, we have made use of the equations of motion of the fields. 

Finally, we consider the invariance under the Lorentz group, i.e., boosts and rotations. The representation of the Lorentz group is the source of spin in quantum field theory, it is then expected that the conserved currents in this case are sensitive to spin polarization. In general, for an infinitesimal Lorentz transformation one has $\delta x_\mu = \omega_{\mu\nu}x^\nu$ with constant  $\omega_{\mu\nu}=-\omega_{\nu\mu}$. The fields will change, according to their representation, following the rule $\delta \phi^a +\delta x^\mu\partial_\mu\phi^a = -i/2\omega_{\mu\nu} (\Sigma^{\mu\nu})^a_b\phi^b$. One obtains then the conserved angular-momentum flux density, 
\begin{equation}
 M^{\lambda,\mu\nu}_c = x^\mu T^{\lambda\nu}_c -x^\nu T^{\lambda\mu}_c -i\frac{\partial{\cal L}}{\partial(\partial_\lambda \phi^a)}\left( \Sigma^{\mu\nu} \right)^a_b \phi^b.
 \label{eq:Mclmn}
\end{equation}
We note that the comma between the first index and the last two is used to emphasize  the fact that $M^{\lambda,\mu\nu}_c=-M^{\lambda,\nu\mu}_c$(different orders of the indices and conventions are used by different authors).

The first two terms in (\ref{eq:Mclmn}) depend on the canonical stress-energy tensor already obtained from the space-time translational invariance of the action and represent the orbital part of the angular momentum. The last term in (\ref{eq:Mclmn}) defines the spin contribution to the angular momentum and is called a canonical spin tensor. 
With
\begin{equation}
 \Sigma^{\mu\nu} = \frac{i}{4} \left[\vphantom{\frac{}{}} \gamma^\mu,\gamma^\nu \right],
\end{equation}
the canonical spin tensor reads
\begin{equation}\label{S_c_Dirac}
 {\cal S}_c^{\lambda , \mu\nu}(x) = \frac{i}{8}\bar\Psi(x) \left\{ \gamma^\lambda,\left[\vphantom{\frac{}{}} \gamma^\mu,\gamma^\nu \right] \right\}\Psi(x).
\end{equation}
Using the anticommutation relations $\{\gamma^\mu,\gamma^\nu\}=2g^{\mu\nu}$, it is straightforward to check that ${\cal S}_c^{\lambda , \mu\nu}$ is totally antisymmetric under the exchange of any indices.

Differently from the total charges (i.e., quantities obtained by the volume integrals), the conserved density currents given by the Noether theorem are not uniquely defined. Whatever the conserved current ${\cal Q}^\mu$ is originally derived, if one builds from the fields a tensor ${\cal C}^{\alpha\mu}=-{\cal C}^{\mu\alpha}$, called a superpotential, the new current ${\cal Q}^{\prime\mu}={\cal Q}^\mu +\partial_\alpha {\cal C^{\alpha \mu}}$ is equally conserved and provides the same conserved total charge. In the particular case of the angular momentum flux and the stress-energy tensor there is a class of well-known transformations that leave the conserved total charges invariant (i.e., the generators of the Poincar\'e group). They are called pseudo-gauge transformations and have the form~\cite{Hehl:1976vr}
\begin{equation}\label{pseudogauge}
 \begin{split}
  &T^{\prime\mu\nu} = T^{\mu\nu} + \frac{1}{2} \partial_\lambda \left( \vphantom{\frac{}{}} {\cal G}^{\lambda,\mu\nu} -{\cal G}^{\mu,\lambda\nu}-{\cal G}^{\nu,\lambda\mu}  \right), \\
  &{\cal S}^{\prime\lambda,\mu\nu} = {\cal S}^{\lambda,\mu\nu} -{\cal G}^{\lambda,\mu\nu} -\partial_\alpha \Xi ^{\alpha\lambda,\mu\nu}.
 \end{split}
\end{equation}
The tensors $T^{\mu\nu}$ and ${\cal S}^{\lambda,\mu\nu}$ on the right-hand side of Eq.~(\ref{pseudogauge}) can be either the canonical ones or the already transformed ones. The auxiliary tensor ${\cal G}^{\lambda,\mu\nu}$ must be antisymmetric in the last two indices, while $\Xi^{\alpha\lambda,\mu\nu}$ should be antisymmetric in both the first two and the last two indices.

For any pair of the stress-energy and spin tensors, the following relations hold
\begin{equation}\label{conservation_laws}
 \begin{split}
  & \partial_\mu T^{\mu\nu}= 0, \\
  & \partial_\lambda {\cal S}^{\lambda,\mu\nu} = -\left(\vphantom{\frac{}{}} T^{\mu\nu} -T^{\nu\mu} \right),
\end{split}
\end{equation}
and the total four-momentum $P^\mu$ and angular momentum $J^{\mu\nu}$ read

\begin{equation}
 \begin{split}
  & P^\mu = \int d^3x \; T^{0\mu}, \\
  & J^{\mu\nu} = \int d^3 x\,  \left( \vphantom{\frac{}{}} x^\mu T^{0\nu} -x^\nu T^{0\mu} + {\cal S}^{0\mu\nu} \right).
 \end{split}
\end{equation}
By construction the last integrals are equal to those obtained with the canonical tensors, therefore Eqs.~(\ref{conservation_laws}) can be equally well considered as the local four-momentum and angular momentum conservation equations. In this work we are not going to discuss which pair is the most appropriate or convenient to represent the physical densities of the (angular) momentum of a physical system. This point is reviewed in Ref.~\cite{Florkowski:2018fap}.

A very special case of the transformations defined by Eqs.~(\ref{pseudogauge}) is the Belinfante symmetrization procedure. In this case, one starts with the canonical tensors $T_c^{\mu\nu}$ and ${\cal S}^{\lambda,\mu\nu}_c$, and takes ${\cal G}^{\lambda,\mu\nu} = {\cal S}^{\lambda,\mu\nu}_c$ and ${\Xi}^{\alpha\lambda,\mu\nu}=0$. As a result, one obtains a vanishing new spin tensor ${\cal S}^{\lambda,\mu\nu}_B=0$, and the angular momentum conservation becomes just the requirement that the antisymmetric part of $T_B^{\mu\nu}$ vanishes. There is an apparent paradox here, namely, that one starts with ten independent equations for 16+24=40 degrees of freedom in~(\ref{conservation_laws})\footnote{This is so in the case of an arbitrary original spin tensor which is antisymmetric only in the last two indices. For the canonical spin tensor that is totally antisymmetric, the number of independent components is 16+4=20.} and ends up with only four equations for 10 degrees of freedom; the vanishing divergence of a symmetric rank two tensor.

It is possible to resolve this paradox by writing the result of the Belinfante symmetrization in a less deceitful way, namely
\begin{equation}
 \begin{split}
  & \partial_\mu T_B^{\{\mu\nu\}} = 0, \\
  & T^{[\mu\nu]}_B = 0 \Rightarrow \partial_\lambda {\cal S}_c^{\lambda\mu\nu}= -\left(\vphantom{\frac{}{}} T^{\mu\nu}_c -T^{\nu\mu}_c \right),\\
  &P^\mu = \int d^3x \; T^{\{0\mu\}}_B, \qquad J^{\mu\nu} = \int d^3 x\,  \left( \vphantom{\frac{}{}} x^\mu T^{\{0\nu\}}_B -x^\nu T^{\{0\mu\}}_B \right).
 \end{split}
 \label{eq:Belnew}
\end{equation}
The middle line of Eqs.~(\ref{eq:Belnew}) emphasizes an important point --- although the symmetric part $T^{\{\mu\nu\}}_B$ of the Belinfante symmetrized stress-energy tensor $T_B^{\mu\nu}$ is separately conserved and both the total four-momentum and angular momentum can be expressed through $T^{\{\mu\nu\}}_B$, the requirement that the antisymmetric part of $T^{[\mu\nu]}_B$ vanishes should be treated as a complementary set of equations. Indeed, starting with the canonical tensors obtained for the Dirac field~(\ref{T_c_Dirac}) and~(\ref{S_c_Dirac}), and performing the Belinfante symmetrization, one obtains
\begin{equation}
 T_B^{\mu\nu} =\frac{i}{2}\bar \Psi(x)\gamma^\mu\stackrel{\leftrightarrow}{\partial^\nu} \Psi(x) -\frac{i}{16}\partial_\lambda\left(\Psi(x) \left\{ \gamma^\lambda,\left[\vphantom{\frac{}{}} \gamma^\mu,\gamma^\nu \right] \right\}\Psi(x) \right),
 \label{eq:Tbelnew}
\end{equation}
a formula which is not manifestly symmetric under a $\mu\leftrightarrow\nu$ exchange. In order to show that (\ref{eq:Tbelnew}) is indeed symmetric, one has two options:

 {\it i)} Solve exactly the Euler-Lagrange equations of motion for the fields (possible for a free field) and directly check the symmetry of (\ref{eq:Tbelnew}).

 {\it ii)} Make use of the angular momentum conservation for the canonical tensors, accepting them as another set of equations.

\noindent Consequently, although one needs a rank two, symmetric, and conserved tensor in order to make a comparison with kinetic theory (since the stress-energy tensor in kinetic theory is symmetric by construction, see Eq~(\ref{kinetic})), one can always consider the equation
\begin{equation}\label{canonical_spin_evolution}
 \partial_\lambda {\cal S}_c^{\lambda\mu\nu} = -\partial_\lambda {\cal S}_c^{\lambda\mu\nu}= -\left(\vphantom{\frac{}{}} T^{\mu\nu}_c -T^{\nu\mu}_c \right),
\end{equation}
which remains valid. We note that the equations for the fields are usually far from being trivial and the same property  holds for the symmetrization procedure that starts from some generic $T^{\mu\nu}$ and  non-vanishing ${\cal S}^{\lambda,\mu\nu}$. For instance, a different Lagrangian density having the same Euler-Lagrange equation of motion for the fields, generally lead to different canonical tensors. In any case all of these tensors lead to the same conserved charges and provide the same number of equations. For a modern and more detailed discussion over the different possible choices of $T^{\mu\nu}$ and ${\cal S}^{\lambda,\mu\nu}$, and their physical consequences we refer to Refs.~\cite{Weickgenannt:2020aaf,Speranza:2020ilk}.

If one excludes the particular case of quantum anomalies\footnote{Neither in a free theory nor in the standard model there are anomalies in the conservation laws for the four-momentum and angular momentum. However, one has to check on a case by case basis if this is so while dealing with a generic quantum field theory.}, very similar arguments to those presented above hold also in the quantum case for the operators built from fundamental fields. In particular, the canonical spin tensor in the quantum case still reads as in Eq.~(\ref{S_c_Dirac}), with the only addition that one has to renormalize it (make normal ordering) to avoid infinities related to the vacuum. The macroscopic, classical, spin tensor is the expectation value of the quantum counterpart. For a generic (pure or mixed) state of the system $\rho$ we have
\begin{equation}\label{macro_spin_tensor}
 {\cal S}_c^{\lambda\mu\nu} = {\rm tr}\left( \vphantom{\frac{}{}}\rho : \hat{\cal S}_c^{\lambda\mu\nu}: \ \right) = \frac{i}{8} {\rm tr}\left( \rho :\bar\Psi(x) \left\{ \gamma^\lambda,\left[\vphantom{\frac{}{}} \gamma^\mu,\gamma^\nu \right] \right\}\Psi(x): \right).
\end{equation}
This form is probably the most intuitive guess for a macroscopic object embedding the particle's polarization degrees of freedom, because in the total angular momentum operator
\begin{equation}\label{hatJ} 
  \hat J^{\mu\nu} = \int d^3 x \; \Psi^\dagger(x)\left( \frac{i}{2}x^\mu \stackrel{\leftrightarrow}{\partial^\nu} -\frac{i}{2}x^\nu \stackrel{\leftrightarrow}{\partial^\mu} +\frac{i}{8}\gamma^0\left \{\gamma^0,\left[ \gamma^\mu,\gamma^\nu\vphantom{\frac{}{}} \right]\right\} \right)\Psi(x),
\end{equation}
 $\hat{\cal S}_c^{0,\mu\nu}$ describes the last term in the round brackets, depending on the gamma matrices. It is the only term that mixes the components of the spinor fields --- the part stemming from $T_c^{\mu\nu}$ depends on the gradients, hence, can be interpreted as the relativistic QFT analogue of ${\bf x}\times{\bf p}$, the orbital angular momentum of classical particles.

In general, similarly to the stress-energy tensor and the current, the macroscopic spin tensor~(\ref{macro_spin_tensor}) depends on both space-time coordinates and two momentum variables. However, its volume integral can be written in terms of a single momentum variable and the expectation values of creation/destruction operators, much like we have seen in the previous section. The space integral of ${\cal S}_c^{0ij}$ can be expected to be related to the sum of the spin polarizations of particles. Defining the vector of matrices $ \Sigma_i $ in the following way
\begin{equation}
  \Sigma_i = \frac{i}{4} \varepsilon_{ijk} [\gamma^j,\gamma^k] = \left( 
 \begin{matrix} 
  \sigma_i & 0 \\
  0 & \sigma_i
 \end{matrix}\right), \qquad \forall i\in\{1,2,3\}
\end{equation}
with $\sigma_i$ being the $2\times 2$ Pauli matrices, one has
\begin{equation}\label{total_spin_c}
 \begin{split}
  & \frac{1}{2}\varepsilon_{ijk} \int \!\!\!d^3x \;{\cal S}_c^{0jk}  \\
  & = \frac{1}{2} \sum_{r,s}\int \!\!\! d^3x\int \frac{d^3p d^3p^\prime}{(2\pi)^6\sqrt{2E_{\bf p}2E_{{\bf p}^\prime}}}\left[ \vphantom{\frac{}{}} \langle a^\dagger_r({\bf p})a_s({\bf p}^\prime)\rangle  U^\dagger_r({\bf p}) \Sigma_i U_s({\bf p}^\prime) e^{i(p-p^\prime)\cdot x} \right.  \\
 &\left. \frac{}{}\qquad \qquad\qquad \qquad - \langle b^\dagger_r({\bf p}) b_s({\bf p}^\prime)\rangle  V^\dagger_s({\bf p}^\prime) \Sigma_i V_r({\bf p}) e^{i(p-p^\prime)\cdot x}  \right.\\
&\left. \frac{}{}\qquad \qquad\qquad \qquad + \langle a^\dagger_r({\bf p}) b^\dagger_s({\bf p}^\prime)\rangle  U^\dagger_r({\bf p}) \Sigma_i V_s({\bf p}^\prime) e^{i(p+p^\prime)\cdot x}  \right. \\
&\left. \frac{}{}\qquad \qquad\qquad \qquad  + \langle b_s({\bf p}) a_r({\bf p}^\prime)\rangle  V^\dagger_s({\bf p}) \Sigma_i U_r({\bf p}^\prime) e^{-i(p+p^\prime)\cdot x}  \right]  \\\\
\end{split}
\end{equation}
that can be rewritten as
\begin{equation}\label{total_spin_c2}
 \begin{split}
 & \frac{1}{2} \sum_{r,s} \int \frac{d^3p}{(2\pi)^3 2E_{\bf p}} \left[ \vphantom{\frac{}{}} \langle a^\dagger_r({\bf p})a_s({\bf p})\rangle  U^\dagger_r({\bf p}) \Sigma_i U_s({\bf p})  \right.  \\
 &\left. \frac{}{}\qquad \qquad\qquad \qquad - \langle b^\dagger_r({\bf p}) b_s({\bf p})\rangle  V^\dagger_s({\bf p}) \Sigma_i V_r({\bf p}) \right.\\
&\left. \frac{}{}\qquad \qquad\qquad \qquad + \langle a^\dagger_r({\bf p}) b^\dagger_s(-{\bf p})\rangle  U^\dagger_r({\bf p}) \Sigma_i V_s(-{\bf p}) e^{2 \, i \, E_{\bf p}\,  t}  \right. \\
&\left. \frac{}{}\qquad \qquad\qquad \qquad  + \langle b_s(-{\bf p}) a_r({\bf p})\rangle  V^\dagger_r(-{\bf p}) \Sigma_i U_s({\bf p}) e^{-2\, i \, E_{\bf p} \, t}  \right] . 
 \end{split}
\end{equation}
Making use of the direct formulas for the massive eigenspinors\footnote{Note that in the Weyl representation of the Clifford algebra a different explicit formula for the massive eigenspinors is typically used, but the general conclusions remain the same.}

\begin{equation}
 \begin{split}
   U_r({\bf p}) &= \sqrt{E_{\bf p}+m}\left(\begin{matrix}
															 & \phi_r \\
															 \frac{{\boldsymbol \sigma} \cdot {\bf p}}{E_{\bf p}+m}&\phi_r
				   										   \end{matrix}\right),\\
   V_r({\bf p}) &= \sqrt{E_{\bf p}+m} \left(\begin{matrix}
															\frac{{\boldsymbol \sigma} \cdot {\bf p}}{E_{\bf p}+m}&\chi_r \\
															&\chi_r
														  \end{matrix}\right),
 \end{split}
\end{equation}
with the two component vectors $\phi_r$ and $\chi_r$ being the eigenstates of the matrix $\sigma_z={\rm diag}(1,-1)$,

\begin{equation}
 \begin{split}
  \phi_1 &= \left(\begin{matrix} 1 \\0 \end{matrix}\right), \qquad  \phi_1 = \left(\begin{matrix} 0 \\1 \end{matrix}\right), \\
  \chi_1 &= \left(\begin{matrix} 0 \\1 \end{matrix}\right), \quad  \chi_2 = -\left(\begin{matrix} 1 \\0 \end{matrix}\right),
 \end{split}
\end{equation}
and the standard relations between the Pauli matrices

\begin{equation}
 \{\sigma_i,\sigma_j\} = 2\delta_{i,j}, \qquad [\sigma_i,\sigma_j] = 2 i \, \varepsilon_{ijk}\, \sigma_k,
\end{equation}
one can rewrite the matrix elements in~(\ref{total_spin_c2}) in the following way

\begin{equation}
 \begin{split}
  U^\dagger_r({\bf p}) \Sigma_i U_s({\bf p}) &= 2m \;\phi_r\sigma_i\phi_s +\frac{2p_i}{E_{\bf p} +m} \phi_r({\bf p}\cdot{\boldsymbol \sigma})\phi_s, \\
  V^\dagger_s({\bf p}) \Sigma_i V_r({\bf p}) &= 2m \;\chi_s\sigma_i\chi_r +\frac{2p_i}{E_{\bf p} +m} \chi_s({\bf p}\cdot{\boldsymbol \sigma})\chi_r, \\
  U^\dagger_r({\bf p}) \Sigma_i V_s(-{\bf p}) &= \left( V^\dagger_s(-{\bf p}) \Sigma_i U_r({\bf p})  \right)^* = -2 i \sum_{j,k} \varepsilon_{ijk} \, p_j\,  \phi_r \sigma_k \chi_s.
 \end{split}
\end{equation}
The first two terms do not correspond exactly to the polarization in the $i$'th direction of a particle or an antiparticle in an eigenstate of four-momentum $p^\mu$, but they are very closely linked to it (we discuss this point in more detail in the next section). In any case, the volume integral of the canonical spin tensor is strongly related to the polarization state of the fundamental excitations of the fields, before any phenomenological approximation. Because of this, it is a reasonably good candidate to study, if one wants to extend the standard treatment of hydrodynamics to include polarization degrees of freedom. 

In general, the awkward mixed terms involving creation and destruction operators are present in Eq.~(\ref{total_spin_c}). They introduce time dependence in the volume integral~(\ref{total_spin_c}) and have no clear interpretation in terms of particle-antiparticle degrees of freedom. Regarding the time dependence, this is not completely unexpected since the canonical spin tensor is not conserved, see Eq.~(\ref{canonical_spin_evolution}), hence the volume integral depends on the time at which it is done. As explained in Appendix~\ref{app:density}, the expectation values of the mixed terms $\langle a^\dagger b^\dagger\rangle$ and $\langle a b\rangle $ do not vanish only if the quantum state of the system is a superposition of states, and among them, some states differ in the number of particles/antiparticles by a single particle-antiparticle pair. How much relevant are these kind of states in a heavy-ion collision environment is yet to be understood.

If one considers non-canonical spin tensors obtained through a pseudo-gauge transformation~(\ref{pseudogauge}), it is possible to remove the time dependent part. For instance, the transformation proposed in~\cite{de_Groot} has the form
\begin{equation}
 \hat \Xi^{\alpha\beta,\mu\nu}=0, \qquad \hat {\cal G}^{\lambda,\mu\nu} =- \frac{1}{8m}\bar \Psi(x)\left( \left[ \vphantom{\frac{}{}}\gamma^\lambda,\gamma^\mu \right] \stackrel{\leftrightarrow}{\partial^\nu} - \left[ \vphantom{\frac{}{}}\gamma^\lambda,\gamma^\nu \right] \stackrel{\leftrightarrow}{\partial^\mu} \right)\Psi(x),
\end{equation}
which results in a conserved spin tensor
\begin{equation}
 {\cal S}^{\lambda,\mu\nu} = {\cal S}_c^{\lambda\mu\nu} - {\cal G}^{\lambda,\mu\nu}, \qquad \partial_\lambda {\cal S}^{\lambda,\mu\nu} =0.
\end{equation}

Hence the flux of the spin density, described by ${\cal S}^{\lambda,\mu\nu}$, across the freeze-out hypersurface is equal to the volume integral of ${\cal S}^{0,\mu\nu}$ at later times. The latter can be computed following the same steps used for the canonical tensor. After  lengthy but straightforward calculations one obtains a time independent formula that is still strongly related to the polarization degrees of freedom,
\begin{equation}\label{total_spin_de_Groot}
 \begin{split}
  \!\!\!& \!\!\!\frac{1}{2}\varepsilon_{ijk} \int \!\!\!d^3x \;{\cal S}^{0,jk} = \\
 \!\!\!& \!\!\!= \frac{1}{2} \sum_{r,s} \int \frac{d^3p}{(2\pi)^3  } \left[  \langle a^\dagger_r({\bf p}) a_s({\bf p})\rangle \left( \frac{E_{\bf p}}{m} \, \phi_r\sigma_i\phi_s - \frac{ p^i}{m(E_{\bf p} + m)} \phi_r ({\bf p}\cdot{\boldsymbol \sigma}) \phi_s\right)  \right.  \\
 &\left.  - \langle b^\dagger_r({\bf p}) b_s({\bf p})\rangle \left( \frac{E_{\bf p}}{m} \, \chi_s\sigma_i\chi_r - \frac{p^i}{m(E_{\bf p} + m)} \chi_s ({\bf p}\cdot{\boldsymbol \sigma}) \chi_r\right)  \right] . 
 \end{split}
\end{equation}
It is important to note that the term in the brackets, despite reducing to the polarization of a two component spinor in the non relativistic limit ($m\to \infty$), does not correspond to the polarization of a relativistic particle. Therefore the last integral must not be confused with the the average (relativistic) polarization multiplied by the average number of (anti)particles. For a discussion of the spin tensor in the context of relativistic thermodynamics we refer to~\cite{Becattini:2013fla,Becattini:2018duy}, see  also Becattini's review in this monograph~\cite{Becattini:2020sww}.

\section{Particle polarization, the Wigner distribution, and the polarization flux pseudotensor}

We have already shown that the spin tensor is a macroscopic object sensitive to the polarization of the excitations of a free quantum field. In this section, we show that the relativistic Wigner distribution is the appropriate extension of the distribution function, that takes into account the polarization degrees of freedom. In particular, the Wigner distribution of a generic state of the system can be linked to the average polarization of particles with fixed momentum, which is probably the most important thing from the point of view of comparisons of theory predictions with experimental data.

Our starting point is a relativistic polarization pseudovector, namely, the relativistic counterpart of the expectation value $\langle\psi|{\boldsymbol \sigma}|\psi\rangle$ used for a two-component, non-relativistic spinor. An important property to take into account is that the latter is a constant of motion for free particles (the free hamiltonian commutes with the Pauli matrices). Thus, the most straightforward way to generalize the concept of $\langle\psi|{\boldsymbol \sigma}|\psi\rangle$ is to look at the classical (non-quantum) relativistic generalization of the internal angular momentum and to apply the same reasoning to the operators in QFT.

For a classical (extended) object the angular momentum reads ${\bf j} ={\bf x}\times{\bf p} + {\bf s}$, with ${\bf s}$ being the intrinsic angular momentum\footnote{At the classical level, the latter corresponds to rotation with respect to an internal axis of the extended object}. The immediate relativistic generalization is~\cite{Mathisson,Weyssenhoff}
\begin{equation}\label{rel_j_class}
 j^{\mu\nu}= x^\mu p^\nu -x^\nu p^\mu + s^{\mu\nu},
\end{equation}
with an antisymmetric tensor $s^{\mu\nu}=-s^{\nu\mu}$. It is easy to notice that the components $(1/2)\sum_{j,k}\varepsilon_{ijk}j^{jk}$ describe the angular momentum, while the components $j^{0i}$ are needed for relativistic covariance, to have the correct transformation rules changing the frame of reference. 

The polarization pseudovector $\Pi^\mu$ is proportional to the dual of the angular momentum, contracted with the four-momentum\footnote{A very similar definition is used for the Pauli-Lubanski pseudovector. It follows the same construction procedure, but without mass in the denominator. Beside different physical dimensions, it is a very close concept which is well defined in the massless case. Since we focus on massive fields herein, we are not going to analyze it. It is useful to notice, however, that using the Pauli-Lubanski definition one can follow the same steps in the massless case, obtaining the helicity distribution instead of the polarization one.}
\begin{equation}\label{polarization_class}
 \Pi^\mu = -\frac{1}{2m}\varepsilon^{\mu\nu\rho\sigma}j_{\nu\rho}p_\sigma =  -\frac{1}{2m}\varepsilon^{\mu\nu\rho\sigma}s_{\nu\rho} p_\sigma.
\end{equation}
Here we have made use of the definition~(\ref{rel_j_class}), removing the contribution of the orbital momentum since it is proportional to the four-momentum and vanishes after contraction with the Levi-Civita symbol. In any inertial reference frame comoving with the system (hence for ${\bf p}=0$) the polarization pseudovector has a vanishing time component, and the space components are just the intrinsic angular momentum ${\bf s}$.

Since $\Pi^{\mu}$ is the contraction of an antisymmetric object with the totally antisymmetric $\varepsilon^{\mu\nu\rho\sigma}$ and the four-momentum $p^\mu$, one needs only tree components to fully describe it, for instance, the space ones. Having these comments in mind, we obtain
\begin{equation}\label{Pi_0}
 0=p_\mu \Pi^\mu= E_{\bf p} \Pi^0 -{\bf p}\cdot {\boldsymbol \Pi} \Rightarrow \Pi^0 = \frac{{\bf p}\cdot {\boldsymbol \Pi} }{E_{\bf p}}.
\end{equation}
It is particularly useful to write the polarization pseudovector in the comoving frame, ${\boldsymbol \Pi}_{\rm com.}$, in terms of the polarization in the lab frame. It will serve us later to make a direct connection between the relativistic polarization operator and the non relativistic one, $\langle\psi|{\boldsymbol \sigma}|\psi\rangle$.

A boost $\Lambda_{\bf p}$ from the lab frame to the comoving frame is characterized by the speed $\beta = \|{\bf p}\|/E_{\bf p}$ and the Lorentz gamma factor $\gamma=1/\sqrt{1-\beta^2}= E_{\bf p}/m$. The zeroth component of $\Pi^{\mu}$ must vanish after such a boost, as immediately follows from Eq.~(\ref{Pi_0}),
\begin{equation}
 \Pi^0\stackrel{\Lambda_{\bf p}}{\longrightarrow} \Pi^0_{\rm com.} =\gamma\left( \Pi^0 -\beta \frac{{\boldsymbol\Pi}\cdot{\bf p}}{\|{\bf p}\|} \right) = \gamma\left( \Pi^0 - \frac{{\boldsymbol\Pi}\cdot{\bf p}}{E_{\bf p}} \right) \equiv 0.
\end{equation}
On the other hand the non-trivial spatial part reads
\begin{equation}\label{com_pol}
 \begin{split}
  {\boldsymbol \Pi}_{\rm com.} &= {\boldsymbol \Pi} -\frac{ {\boldsymbol \Pi}\cdot{\bf p}}{\|{\bf p}\|^2}{\bf p} + \gamma\left( \frac{ {\boldsymbol \Pi}\cdot{\bf p}}{\|{\bf p}\|} -\beta \Pi^0 \right)\frac{{\bf p}}{\|{\bf p}\|}  \\
  &= {\boldsymbol \Pi} -\frac{ {\boldsymbol \Pi}\cdot{\bf p}}{\|{\bf p}\|^2}\left[ 1 -\frac{E_{\bf p}}{m}\left(1 - \frac{\|p\|^2}{E_{\bf p}^2}  \right) \right]{\bf p} \\
  &= {\boldsymbol \Pi} -\frac{ {\boldsymbol \Pi}\cdot{\bf p}}{\|{\bf p}\|^2} \left[ \frac{E_{\bf p}-m}{E_{\bf p}}\equiv \frac{\|{\bf p}^2\|}{E_{\bf p}(E_{\bf p}+m)} \right]{\bf p} \\
  &= {\boldsymbol \Pi} -\frac{ {\boldsymbol \Pi}\cdot{\bf p}}{E_{\bf p}(E_{\bf p}+m)}{\bf p}.
 \end{split}
\end{equation}
In relativistic quantum field theory one has the operator analogue of the polarization~(\ref{polarization_class}) for a massive Dirac field, namely, the operator
\begin{equation}
 \hat \Pi^\mu = -\frac{1}{2m} \varepsilon^{\mu\nu\rho\sigma}:\hat J_{\nu\rho}::\hat P_\sigma:\,.
\end{equation}
We note that one has to use normal ordering for the two operators separately, because otherwise  the expectation value of $\hat \Pi^\mu$ in single-particle states would vanish\footnote{If one applies the normal ordering $:\hat J_{\nu\rho}\hat P_\sigma:$ at the operator level there are two destruction operators on the left-hand side, which annihilate any single particle state. One would need at least two (anti)particle states to have a non-vanishing expectation value.}. The most general one-particle state $|\psi_1\rangle$ for a free field reads
\begin{equation}
 |\psi_1\rangle = \sum_r\int\frac{d^3 p}{(2\pi)^3 2 E_{\bf p}}\psi_1(p,r)|p,r\rangle,
\end{equation}
with the normalization
\begin{equation}\label{wave_norm}
 1=\langle\psi_1|\psi_1\rangle = \sum_r \int\frac{d^3 p}{(2\pi)^3 2E_{\bf p}}\psi_1^*(p,r)\psi_1(p,r),
\end{equation}
in which we made use of the normalization of the states
\begin{equation}
 \langle p,r|q,s\rangle = 2E_{\bf p} (2 \pi)^3 \delta_{rs}\,  \delta^3({\bf p}-{\bf p}^\prime).
\end{equation}
The polarization vector then reads
\begin{equation}\label{pol_def}
 \begin{split}
  \langle\psi_1|\hat \Pi^\mu|\psi_1\rangle& =-\frac{1}{2m}\varepsilon^{\mu\nu\rho\sigma}\sum_{r,r^\prime} \int\frac{d^3 p d^3p^\prime}{(2\pi)^6 2E_{\bf p} 2E_{{\bf p}^\prime}}\psi_1^*(p^\prime,r^\prime)\psi_1(p,r) \times\\
  & \qquad \qquad \times \langle p^\prime,r^\prime|:\hat J_{\nu\rho}::\hat P_\sigma:|p,r\rangle.
 \end{split}
\end{equation}
Using the anticommutation relations
\begin{equation}
 \{a_s({\bf q}),a^\dagger_r({\bf p})\} = (2\pi)^3 \delta_{rs}\delta^3({\bf p}-{\bf q}),
\end{equation}
and taking into account the definition
\begin{equation}
 |p,r\rangle = \sqrt{2E_{\bf p}} a^\dagger_r({\bf p})|0\rangle,
\end{equation}
it is relatively straightforward to prove that
\begin{equation}
 :\hat P_\sigma:|p,r\rangle = p_\sigma|p,r\rangle,
\end{equation}
where $\hat P^\mu$ is the total four-momentum operator,
\begin{equation}
 :\hat P^\mu: = \sum_s \int\frac{d^3 q}{(2\pi)^3} q^\mu\left[ \vphantom{\frac{}{}} a^\dagger_s({\bf q})a_s({\bf q}) + b^\dagger_s({\bf q})b_s(\bf q) \right].
\end{equation}

The situation is slightly more complicated for the expectation value of the angular momentum operator $\langle p^\prime,r^\prime|:\hat J_{\nu\rho}:|p,r\rangle$. 
One can check that a non-zero contribution reads
\begin{equation}\label{pol_1}
 \begin{split}
  &\langle\psi_1|\hat \Pi^\mu|\psi_1\rangle =-\frac{1}{2m}\varepsilon^{\mu\nu\rho\sigma}\sum_{r,r^\prime} \int\frac{d^3 p d^3p^\prime}{(2\pi)^6 2E_{\bf p} 2E_{{\bf p}^\prime}}\psi_1^*(p^\prime,r^\prime)\psi_1(p,r) p_\sigma \times\\
  & \qquad \qquad \qquad  \times \langle p^\prime,r^\prime|:\hat J_{\nu\rho}:|p,r\rangle= \\
  & = -\frac{1}{2m}\varepsilon^{\mu\nu\rho\sigma}\sum_{r,r^\prime}\int\!\!\!d^3 x \int\frac{d^3 p d^3p^\prime}{(2\pi)^6 2E_{\bf p} 2E_{{\bf p}^\prime}}\psi_1^*(p^\prime,r^\prime)\psi_1(p,r) p_\sigma \times \\
  &\times \frac{i}{8}U^\dagger_{r^\prime}({\bf p}^\prime)\gamma^0\left \{\gamma^0,\left[ \gamma_\nu,\gamma_\rho\vphantom{\frac{}{}} \right]\right\}U_r({\bf p}) e^{-i(p-p^\prime)\cdot x}= \\
  &=  -\frac{1}{4m}\varepsilon^{\mu i j \sigma}\varepsilon_{ijk}\sum_{r,r^\prime} \int\frac{d^3 p}{(2\pi)^3 }\frac{\psi_1^*(p,r^\prime)\psi_1(p,r)}{2E_{\bf p}}\frac{ p_\sigma }{2E_{\bf p}} U^\dagger_{r^\prime}({\bf p})\Sigma_k U_r(\bf p).
 \end{split}
\end{equation}
In particular, the space part of the polarization $\langle\psi_1|\hat {\boldsymbol \Pi}|\psi_1\rangle$ reads
\begin{equation}
 \begin{split}
  \langle\psi_1|\hat {\boldsymbol \Pi}|\psi_1\rangle &= \frac{1}{4m} \sum_{r,s}  \int\frac{d^3 p}{(2\pi)^3 }\frac{\psi_1^*(p,r)\psi_1(p,s)}{2E_{\bf p}}  U^\dagger_{r}({\bf p})\left( \begin{matrix} {\boldsymbol \sigma} &0\\0&{\boldsymbol \sigma} \end{matrix} \right) U_s(\bf p) = \\
  & = \frac{1}{2} \sum_{r,s} \int\frac{d^3 p}{(2\pi)^3 }\frac{\psi_1^*(p,r)\psi_1(p,s)}{2E_{\bf p}} \left[  \phi_r{\boldsymbol \sigma}\phi_s  +\frac{\phi_r({\bf p}\cdot{\boldsymbol \sigma})\phi_s}{m(E_{\bf p} +m)} \,{\bf p}  \right].
 \end{split}
\end{equation}

At this point, with the help of~(\ref{com_pol}), it is possible to highlight the link between the expectation value we have just computed and the non-relativistic polarization $\langle\psi|{\boldsymbol \sigma}|\psi\rangle$. We first define the spin momentum-dependent density matrix

\begin{equation}\label{frs_p}
 f_{rs} ({\bf p}) = \frac{\psi_1^*(p,r)\psi_1(p,s)}{2E_{\bf p}},
\end{equation}
which is a two-by-two Hermitian matrix in the indices $r,s$ for every value of the momentum ${\bf p}$ and describes a polarized state. It is normalized to one, i.e., its trace over the $r,s$ indices is unitary while integrated with the measure $\int d^3p/(2\pi)^3$  (because of the normalization of the wave function~(\ref{wave_norm})). Taking into account a momentum eigenstate of the form $f_{rs}=(2\pi)^3 H_{rs} \delta^3({\bf p} -\tilde{\bf p})$~\footnote{This is actually forbidden, since the wave function $\psi_1$ is a regular distribution in momentum. However, one can have the spin density matrix factorized in a Hermitian $2\times 2$ matrix times and arbitrarily sharp gaussian in the momentum. Such a strongly delocalized state is, for all practical purposes, equivalent to a momentum eigenstate.}, one finds the polarization
\begin{equation}
  \frac{1}{2} \sum_{r,s}H_{rs} \left[  \phi_r{\boldsymbol \sigma}\phi_s  +\frac{\phi_r(\tilde{\bf p}\cdot{\boldsymbol \sigma})\phi_s}{m(E_{\tilde{\bf p}} +m)} \,\tilde{\bf p}  \right],
\end{equation}
with $H_{rs}=H^*_{sr}$ and ${\rm tr}(H)=1$. Making use of~(\ref{com_pol}),  one finds that the polarization in the comoving frame reads

\begin{equation}
 \frac{1}{2} \sum_{r,s}H_{rs} \phi_r{\boldsymbol \sigma}\phi_s,
\end{equation}
which is, indeed, the polarization of a non-relativistic spinor. Hence, it is limited between $-1/2$ and $1/2$ in each direction. 

It is important to note, however, that the polarization in the lab frame is not limited. For instance

\begin{equation}
 H =\left( \begin{matrix} 1 & 0 \\ 0 &0 \end{matrix} \right) \Rightarrow \langle\hat{\boldsymbol \Pi}\rangle=\frac{1}{2}\left( \begin{matrix} \frac{\tilde{p}_z \tilde{p}_x}{m(E_{\tilde{\bf p}} + m)} \\ \frac{\tilde{p}_z \tilde{p}_y}{m(E_{\tilde{\bf p}} + m)} \\ 1+ \frac{\tilde{p}_z^2}{m(E_{\tilde{\bf p}} + m)} \end{matrix} \right),
\end{equation}
which has, manifestly, arbitrarily large components as long as $\tilde{p}_z\neq 0$, maintaining the expected polarization $(0,0,1/2)$ in the comoving frame for a $z$ polarized state. For a more general case, i.e., if $f_{rs}\slashed\propto\delta^3({\bf p} -\tilde{\bf p})$, one must keep the relativistic corrections. 
 
If one considers the defining relation~(\ref{frs_p}), the spin density matrix has an immediate physical interpretation. The trace
\begin{equation}
\sum_r\left[\frac{f_{rr}({\bf p})}{(2\pi)^3}\right]
\end{equation}
is the probability density to obtain ${\bf p}$ in a momentum measurement, while the average polarization in the comoving frame reads
\begin{equation}\label{average_pol}
 \frac{1}{2}\sum_{rs}\frac{\left[{f_{rs}({\bf p})}\left( \phi_r{\boldsymbol \sigma}\phi_s \right)\right]}{\sum_t\left[{f_{tt}({\bf p})}\right]}.
\end{equation}
We thus see that the spin density matrix is sufficient to characterize the most important experimental observables for a free spin $1/2$ particle. It is understood that all the steps can be repeated for a single antiparticle wave-function to obtain the antiparticle polarization, which depends on the antiparticle spin density matrix $\bar f_{rs}({\bf p})$. The only significant difference is an overall $-1$ sign, because of the corresponding sign in the spin part of the normal ordered angular momentum operator, and an exchange of the $r$ and $s$ indices in the $\chi$ bispinors compared to the $\phi$ for the particles~\footnote{Which can be expected, since the conventional two component spinors $\chi$ in the negative frequency solutions of the Dirac equation are taken with the opposite eigenvalue of $\sigma_z$, compared to the positive frequency solutions.}.

The appropriate generalization of the classical distribution function is expected, therefore, to produce in some limit a multi-particle generalization of the spin density matrix that provides both the spectrum in momentum of the produced particles and their average polarization. As we have already anticipated, the desired object is the Wigner distribution. Its most convenient definition is
\begin{equation}\label{Wigner_def}
 \begin{split}
   & \hat W_{AB}(x,k) = \int\!\!\! \frac{d^4 v}{(2\pi)^4} \, e^{-ik\cdot v} \; \Psi^{\dagger }_B (x+v/2) \Psi_A (x-v/2) ,
 \end{split}
\end{equation}
that is, it is a four by four matrix obtained from thecomponents of two fields, $  \Psi(y)^{\dagger}_B\Psi_A(z)$, Fourier transformed with respect to the relative distance $v=z-y$, and with $x=(z+y)/2$ being the middle space-time point. The inversion on the relative position of the matrix elements between the left and right hand side of the last equation is neede in order to use the ordinary rules in matrix multiplications with respect to the $A,B$ indices. In the remander of this work we will omit the matrix indices, understanding the matrix nature, and we will just write, eg, $U_r U^\dagger_r$ without indices in a similar way, understanding the fact that it is a $4\times 4$ matrix.

An important property of the Wigner distribution~(\ref{Wigner_def}) is that it is a Hermitian matrix representing physical observables (at least in principle). Moreover, one expects that the usual causality rules apply to it. It is worth mentioning that some authors use a different sign convention or use  $\bar\Psi_B \Psi_A$ in the definition of $W(x,k)$~\cite{Florkowski:2018ahw,Weickgenannt:2019dks,Liu:2020flb}, thus making the alternatively defined matrix a non-Hermitian one, so one must check which version of the Wigner distribution is actually used while comparing different works. In any case a matrix multiplication with $\gamma^0$ and an eventual multiplication by a constant is enough to switch notation.

By the correspondence principle, the classical distribution is the expectation value of the renormalized operator
\begin{equation}\label{Wigner_macro}
 \begin{split}
  W(x,k) &= {\rm tr}\left( \vphantom{\frac{}{}} :\hat W (x,k): \right).
 \end{split}
\end{equation}
Making use of the definition~(\ref{Wigner_def}), and assuming some minimal smoothness of the integrals\footnote{In order to exchange the order of the integrations and integrate by parts.}, one can rewrite the expectation value of any bilinear form in the Dirac fields using  integration over the momentum $k$ of the trace of the macroscopic Wigner distribution~(\ref{Wigner_macro})
\begin{equation}\label{conversion}
 \begin{split}
  & {\rm tr}\left( \rho : \bar\Psi(x) \gamma^{\nu_1}\cdots\gamma^{\nu_n}\, \frac{i}{2} \stackrel{\leftrightarrow}{\partial^{\mu_1}}\cdots \frac{i}{2} \stackrel{\leftrightarrow}{\partial^{\mu_m}} \Psi(x): \right)  \\
  & \qquad = \int \!\!\! d^4 k \; k^{\mu_1}\cdots k^{\mu_m}\;  {\rm tr}_4\left( \vphantom{\frac{}{}} W(x,k)\gamma^0 \, \gamma^{\nu_1}\cdots \gamma^{\nu_n} \right).
 \end{split}
\end{equation}
Here, the trace on the left-hand side is the usual trace over the quantum states, while ${\rm tr}_4$ on the right-hand side denotes the trace over the matrix indices. To derive the formula  (\ref{conversion}) we have used the integral representation of the Dircac delta $\delta^4(v)=\int d^4k/(2\pi)^4 \exp\{-i k\cdot v\}$ and performed the integration by parts
\begin{equation}\label{k_to_grad}
 \int d^4 v \; k^\mu \,e^{-ik\cdot v}[\cdots] = \int d^4 v \left(i\partial^\mu_{(v)} \,e^{-ik\cdot v}\right) [\cdots] =  \int d^v k  \;e^{-ik\cdot v} \left( -i\partial^\mu_{(v)}\right) [\cdots].
\end{equation}
In the next step, we can use the definition of the Dirac fields to expand~(\ref{Wigner_macro}) and to explicitly represent it in terms of the expectation values of the creation/destruction operators
\begin{equation}\label{Wigner_expand}
 \begin{split}
  W(x,k) &= \sum_{rs}\int\!\!\! \frac{d^4 v}{(2\pi)^4} e^{-ik\cdot v}\int\frac{d^3p d^3 q}{(2\pi)^6 \sqrt{2E_{\bf p}2E_{\bf q}}} \left[\vphantom{\frac{}{}}  \right. \\
  & \qquad \langle a^\dagger_r({\bf p}) a_s({\bf q})\rangle U^\dagger_r({\bf p}) U_s({\bf q}) e^{i(p-q)\cdot x}e^{i\left(\frac{p+q}{2}\right)\cdot v} + \\ \\
 & \quad -\langle b^\dagger_r({\bf p}) b_s({\bf q})\rangle V_s^\dagger({\bf q}) V_r({\bf p}) e^{i(p-q)\cdot x}e^{-i\left(\frac{p+q}{2}\right)\cdot v} + \\ \\
  & \quad + \langle b_s({\bf q})a_r({\bf p})\rangle  V_s^\dagger({\bf q}) U_r({\bf p}) e^{-i(p+q)\cdot x}e^{i\left(\frac{p-q}{2}\right)\cdot v} + \\ \\
  & \quad \left.  \! + \langle a^\dagger_r({\bf p}) b^\dagger_s({\bf q})\rangle  U_r^\dagger({\bf p}) V_s({\bf q}) e^{i(p+q)\cdot x}e^{i\left(\frac{p-q}{2}\right)\cdot v} \right],
 \end{split}
\end{equation}
it is important to remind that both sides must be a matrix, therefore the eigenspinors are not contracted but must be read, eg, $(U^\dagger_r)_ B(U_s)_A$, according to ~(\ref{Wigner_def}). The last formula~(\ref{Wigner_expand}) allows us to make two important observations. The first one is that, after performing the $d^4 v$ integral, each of the four sectors is proportional to the Dirac delta function  $\delta^4(k \pm (p\pm q)/2)$, with a different combination of the $+$ and $-$ signs in each sector. The momenta $p^\mu$ and $q^\mu$ are both on the mass shell but their combination, in general, is not. The pure particle/antiparticle contributions have $\delta^4(k\pm(p+q)/2)$ which can be on shell if and only if $p=q$. The mixed terms, however, include $\delta^4(k\pm(p-q)/2)$ where $(p-q)/2$ is never on shell and always space-like. This is the reason why we call $k^\mu$ a wave number vector, in order not to confuse it with the four-momentum of some particle-like degree of freedom.

The second and possibly the most important thing to notice is that $k^\mu W(x,k)$ is conserved, i.e., $k^\mu \partial_\mu W(x,k)=0$, as one can check directly by applying $k^\mu\partial_\mu$ to the right-hand side of~(\ref{Wigner_expand}) and using the integration by parts in~(\ref{k_to_grad}) to convert the wavenumber vector $k$ into a derivative with respect to $v$. This is a consequence of the Dirac equation for the fields, which implies that the Klein-Gordon equation is satisfied as well. 

Because of the conservation of $k^\mu W(x,k)$, one can use the same mathematical framework as that already used for the conserved fluxes such as $T^{\mu\nu}$ and the classical expression $p^\mu f(x,{\bf p})$ Here we can see the reason of the choice to define the Wigner operator as Hermitian.  Being $k^\mu W(x,k)$ Hermitian too (an observable) is expected to follow causality rules. As a consequence of the Gauss theorem, the flux over the freeze-out hypersurface (or any other surface following the freeze-out) is equal to the volume integral\footnote{The hypothesis of an isolated system is important too. Being the integrand an observable, the flux over the light-cone starting from the spatial boundary of an isolated system must be vanishing. Causality prevents the Wigner distribution to flow out of the light cone, as it would be a superluminal signal transfer, and the hypothesis of an isolated system prevents any signal to flow inside of the light cone. The flux over the light cone is therefore vanishing.}
\begin{equation}\label{Wigner_flux}
 \int \!\!\! d\Sigma_\mu \, k^\mu \, W(x,k) = \int \!\!\! d^3x \, k^0\,  W(x,k).
\end{equation}
The analysis of the volume integral corresponds to a considerable simplification in the treatment of the Wigner distribution, like for the other quantum objects we have seen. Integrating directly ~(\ref{Wigner_expand}) one obtains
\begin{equation}\label{Wigner_int}
 \begin{split}
  \int \!\! d^3x \, k^0\,  W(x,k) &= \sum_{rs}\int\frac{d^3p}{(2\pi)^3} \frac{k^0}{2E_{\bf p}} \left[  \delta^4(k-p)\,  \langle a^\dagger_r({\bf p}) a_s({\bf p})\rangle\, U^\dagger_r({\bf p}) U_s({\bf p}) + \right. \\ 
  &  \quad \left.  -\delta^4(k+p) \, \langle b^\dagger_r({\bf p}) b_s({\bf p})\rangle \, V_s^\dagger({\bf p}) V_r({\bf p})\vphantom{\frac{}{}} \right].
 \end{split}
\end{equation}
The mixed terms vanish exactly, since the volume integral provides a  $\delta^3({\bf p} +{\bf q})$. Therefore, the Dirac function $\delta^4(k\pm(p-q)/2)$ becomes $\delta(k^0) \delta^3({\bf k}\pm{\bf p})$ in this case. The appearance of $k^0$ makes these terms vanishing, as $k^0\delta(k^0) \delta^3({\bf k}\pm{\bf p})\equiv 0$.

The flux of the Wigner distribution~(\ref{Wigner_flux}) has many interesting properties. They can be identified while looking at its explicit form given by~(\ref{Wigner_int}). It includes an on-shell positive-frequency contribution for the particles and a  negative-frequency (with negative momentum) contribution for the antiparticles. With $k$ being on the mass-shell, one can divide by $k^0$ since $\|k^0\|\ge m$ -- something that cannot be done for the full distribution~(\ref{Wigner_expand}). Having in mind the normalization ${\rm tr}_4(U^\dagger_r({\bf p})\cdot U_s({\bf p})) =U^\dagger_r\cdot U_s = 2E_{\bf p} \delta_{rs} = V^\dagger_r\cdot V_s = {\rm tr}_4(V_r^\dagger({\bf p})V_s({\bf p}))$, and the exact relations (see Appendix~\ref{app:density})
\begin{equation}
 \sum_r \frac{\langle a^\dagger_r({\bf p}) a_r({\bf p})\rangle}{(2\pi^3)} = \frac{dN}{d^3p}, \qquad \sum_r \frac{\langle b^\dagger_r({\bf p}) b_r({\bf p})\rangle}{(2\pi^3)} = \frac{d\bar N}{d^3p}, 
\end{equation}
it is immediate to verify that the trace of the flux~(\ref{Wigner_int}) reads
\begin{equation}\label{exact_spectra}
 \begin{split}
  \int\!\!\!d\Sigma_\mu \, k^\mu W(x,k) = \delta(k^0 - E_{\bf k}) \, E_{\bf k}\, \frac{dN}{d^3p}({\bf k}) + \delta(k^0+E_{\bf k}) \,\,  E_{\bf k} \frac{d\bar N}{d^3p}(-{\bf k}).
 \end{split}
\end{equation}
In other words, the positive-frequency contribution to the flux is directly expressed by the (invariant) spectrum of particles with momentum ${\bf k}$; while the negative-frequency contribution is given by the invariant spectrum of antiparticles with momentum $-{\bf k}$.

The structure of~(\ref{Wigner_int}) is quite rich. It does not include the (anti)particle's spectra only but depends on the polarization states. By construction, the expectation values $\langle a^\dagger_r({\bf p}) a_s({\bf p})\rangle$ and $\langle b^\dagger_r({\bf p}) b_s({\bf p})\rangle$ have a very similar structure to the one-particle spin density matrix~(\ref{frs_p}). They are both Hermitian matrices with respect to the indices $r,s$ for all values of ${\bf p}$. Moreover, their diagonal elements $\langle a^\dagger_r({\bf p}) a_r({\bf p})\rangle$ and $\langle b^\dagger_r({\bf p}) b_r({\bf p})\rangle$ are always non-negative. The only difference is the normalization. Instead of being normalized to $1$ they are normalized to the average number of particles $\langle N\rangle$ and antiparticles $\langle \bar N \rangle$
\begin{equation}
 \begin{split}
  \sum_r\int \!\!\! d^3 p \frac{\langle a^\dagger_r({\bf p}) a_r({\bf p})\rangle}{(2\pi^3)} &= \int \!\!\! d^3 p\frac{dN}{d^3p} = \langle N\rangle, \\
  \sum_r \int \!\!\! d^3 p\frac{\langle b^\dagger_r({\bf p}) b_r({\bf p})\rangle}{(2\pi^3)} &= \int \!\!\! d^3 p\frac{d\bar N}{d^3p} = \langle \bar N\rangle.
 \end{split} 
\end{equation}
They provide therefore the desired generalization of the one-particle spin density matrix to the multi-particle case. The flux of the Wigner distribution~(\ref{Wigner_flux}) directly depends on them, it is therefore not surprising that one can get the average polarization density in momentum space from it. Making use of both~(\ref{Wigner_flux}) and~(\ref{Wigner_int}) we find
\begin{equation} 
 \begin{split}
  &\frac{1}{2m}{\rm tr}_4 \left[ \left(\int\!\!\! d\Sigma_\mu \, k^\mu \, W(x,k) \vphantom{\frac{}{}}\right)\gamma^0\gamma^i\gamma_5 \right] = \\ \\
  & =\frac{1}{4m}\sum_{r,s} \int\!\!\!d^3 p\left[ \delta^4(k-p) \frac{\langle a^\dagger_r({\bf p}) a_s({\bf p})\rangle}{(2\pi^3)}U^\dagger_r({\bf p})\left( \begin{matrix} \sigma_i & 0 \\ 0& \sigma_i \end{matrix} \right)U_s({\bf p}) +\right. \\
  & \qquad \qquad \left. \delta^4(k+p) \frac{\langle b^\dagger_r({\bf p}) b_s({\bf p})\rangle}{(2\pi^3)}V^\dagger_s({\bf p})\left( \begin{matrix} \sigma_i & 0 \\ 0& \sigma_i \end{matrix} \right)V_r({\bf p})\right] =\\ \\
  &= \frac{1}{2}\sum_{r,s}\left\{ \delta(k^0-E_{\bf k}) \frac{\langle a^\dagger_r({\bf k}) a_s({\bf k})\rangle}{(2\pi^3)} \left[  \phi_r{\sigma_i}\phi_s  +\frac{\phi_r({\bf k}\cdot{\boldsymbol \sigma})\phi_s}{m(E_{\bf k} +m)} \, k_i  \right] + \right. \\
  & \qquad \qquad \left. \delta(k^0 + E_{\bf k}) \frac{\langle b^\dagger_r(-{\bf k}) b_s(-{\bf k})\rangle}{(2\pi^3)} \left[  \chi_s{\sigma_i}\chi_r  +\frac{\chi_s({\bf k}\cdot{\boldsymbol \sigma})\chi_r}{m(E_{\bf k} +m)} \, k_i  \right] \right\},
 \end{split}
\end{equation}
which can be immediately recognized as the average polarization of particles with momentum ${\bf k}$ (multiplied by the (non-invariant) spectrum $dN/d^3p({\bf k})$ for the positive frequency) minus the average polarization of antiparticles of momentum $-{\bf k}$ (times the spectrum $d \bar N/d^3p (-{\bf k})$). Since the spectra can be calculated from the flux of the Wigner distribution, one can obtain the average polarizations of particles, $\langle{\boldsymbol \Pi}({\bf p})\rangle$, and antiparticles, $\langle\bar{\boldsymbol \Pi}({\bf p})\rangle$, for any momentum ${\bf p}$,
\begin{equation} 
 \begin{split}
 \langle {\boldsymbol \Pi}({\bf p})\rangle&= \frac{1}{2}\sum_{r,s} \frac{\langle a^\dagger_r({\bf p}) a_s({\bf p})\rangle}{\sum_t\langle a^\dagger_t({\bf p}) a_t({\bf p})\rangle} \left[  \phi_r{\boldsymbol\sigma}\phi_s  +\frac{\phi_r({\bf p}\cdot{\boldsymbol \sigma})\phi_s}{m(E_{\bf p} +m)} \, {\bf p}  \right] , \\ \\
 \langle\bar{\boldsymbol \Pi}({\bf p})\rangle&= -\frac{1}{2}\sum_{r,s} \frac{\langle b^\dagger_s({\bf p}) b_r({\bf p})\rangle}{\sum_t \langle b^\dagger_t({\bf p}) b_t({\bf p})\rangle} \left[  \chi_r{\boldsymbol\sigma}\chi_s  +\frac{\chi_r({\bf p}\cdot{\boldsymbol \sigma})\chi_s}{m(E_{\bf p} +m)} \,{\bf p}  \right].
 \end{split}
\end{equation}
Making use of~(\ref{com_pol}) one can compute the polarization in the comoving frame as usual.

It is straightforward to check that the trace ${\rm tr}_4\left[\frac{1}{2m} \left(\int\!\!\! d\Sigma_\mu \, k^\mu \, W(x,k) \vphantom{\frac{}{}}\right)\gamma^0\gamma^0\gamma_5 \right]$ corresponds to the momentum density of $\Pi^0$ ( i.e., it is equal to the time component of the polarization vector -- particle contribution for the positive frequency minus the antiparticle contribution for the negative frequency). The same structure of the trace is obtained with $\gamma^0$ replaced by $\gamma^i$. One can summarize all these results by making use of the definitions
\begin{equation}
 \begin{split}
  \langle\Pi^0({\bf p})\rangle &= \frac{\langle{\boldsymbol\Pi}({\bf p})\rangle\cdot{\bf p}}{E_{\bf p}}, \qquad \langle\bar\Pi^0({\bf p})\rangle= \frac{\langle{\bar{\boldsymbol\Pi}}({\bf p})\rangle\cdot{\bf p}}{E_{\bf p}}, 
 \end{split}
\end{equation}
to complete the covariant $\langle\Pi^\mu({\bf p})\rangle$ with the correct time component\footnote{Compare with Eq.~(\ref{Pi_0}) for a quick check.}. Thus, the compact form of the previous results on the average polarization reads
\begin{equation}\label{Wigner_pol}
 \begin{split}
  &\frac{1}{2m}{\rm tr}_4 \left[ \left(\int\!\!\! d\Sigma_\lambda \, k^\lambda \, W(x,k) \vphantom{\frac{}{}}\right)\gamma^0\gamma^\mu\gamma_5 \right] =  \\
  &= \delta^4(k^0-E_{\bf k}) \frac{dN}{d^3p}({\bf k})\; \langle \Pi^\mu ({\bf k})\rangle - \delta^4(k^0 + E_{\bf k}) \frac{dN}{d^3p}(-{\bf k})\; \langle \bar \Pi^\mu (-{\bf k})\rangle.
 \end{split}
\end{equation}
The last equation, in conjunction with the exact result in~(\ref{exact_spectra}), is enough to grant that the Wigner at the freeze-out hypersurface is sufficient to predict all the relevant experimental spectra of produced (anti)particles.

In the last section we have seen that the spin tensor is a macroscopic observable sensitive to the microscopic polarization states. Looking at the relaitively simple trace on the left-hand side of~(\ref{Wigner_pol}), one may think if there is macroscopic object related to this. Taking into account the exact conversion rules~(\ref{conversion}), one finds that the $d^4 k $ integral of the left-hand side reads 
\begin{equation}
 \begin{split}
  &\frac{1}{2m}\int \!\!\! d^4 k\, {\rm tr}_4 \left[ \left(\int\!\!\! d\Sigma_\lambda \, k^\lambda \, W(x,k) \vphantom{\frac{}{}}\right)\gamma^0\gamma^\mu\gamma_5 \right] = \\
  & \qquad = \int\!\!\! d\Sigma_\lambda \langle: \frac{i}{4m}\bar\Psi \left(\vphantom{\frac{}{}}\stackrel{\leftrightarrow}{\partial^\lambda} \gamma^\mu\gamma_5\right) \Psi :\rangle.
 \end{split}
\end{equation}
It is, therefore, the flux of (the expectation value of) the rank $2$ pseudotensor
\begin{equation}
 \frac{i}{4m}\bar\Psi(x) \left(\vphantom{\frac{}{}}\stackrel{\leftrightarrow}{\partial^\lambda} \gamma^\mu\gamma_5\right) \Psi(x),
\end{equation}
which we may call the polarization flux pseudotensor, given its relation to the integral of the polarization pseudovector density. It is straightforward to check that its divergence in the $\lambda$ index  vanishes. Hence, it is conserved, as one could expect since its flux is time-independent. It does not correspond to any spin tensor, 
but it provides a valid alternative  as a macroscopic object sensitive to the micorscopic polarization states.

\section{Summary}
\label{sec:conc}

In this work, we have extended the standard kinetic-theory formalism to include spin polarization for particles with spin 1/2. This has been achieved by using the spin tensor and the Wigner function. Our results can be used for the interpretation of the heavy-ion data describing spin polarization of the emitted hadrons.

\begin{acknowledgement}
We would like to thank F.~Becattini, B.~Friman, R.~Ryblewski, and E.~Speranza for insightful discussions. L.T. was supported by the Deutsche Forschungsgemeinschaft (DFG, German Research Foundation) through the CRC-TR 211 ``Strong-interaction matter under extreme conditions'' - project number 315477589 – TRR 211. W.F. was supported in part by the Polish National Science Center Grants No.~2016/23/B/ST2/00717.
\end{acknowledgement}
\backmatter
\appendix

\chapter{Expectation values of creation and destruction operators}
\label{app:density}

In this appendix we show details of the calculations of the expectation values of creation and destruction operators. In particular, we find an interesting and  intuitive link between the average (anti)particle number and the quantum fluctuations required for the mixed terms ($\langle a^\dagger b^\dagger\rangle$ and $\langle a b\rangle$) to be non-vanishing.

The starting point is the density matrix~(\ref{density_matrix}), which reads
\begin{equation}\label{density}
 \rho = \sum_i {\sf P}_i \left| \psi_i \right\rangle \left\langle  \psi_i\right|.
\end{equation}
All ${\sf P}_i$'s are classical probabilities
\begin{equation}
 \sum_i {\sf P}_i =1.
\end{equation}
The states $|\psi_i\rangle$ are proper quantum states, that is, they are normalized to one
\begin{equation}
 \langle\psi_i|\psi_i\rangle =1, \qquad \forall i.
\end{equation}
The expectation value ${\cal O}$ of any quantum operator\footnote{In general one needs the renormalized operators. For free fields this is just the normal ordering, that is, removing the vacuum expectation value. We always assume massive free Dirac fields and normal ordering in this section.} $\hat {\cal O}$ is a weighted average of the expectation values in the pure states, with the classical weights ${\sf P}_i$,
\begin{equation}
 {\cal O} = {\rm tr}\left( \vphantom{\frac{}{}} \rho \, \hat{\cal O} \right) = \sum_i {\sf P}_i \, {\rm tr}\left( \vphantom{\frac{}{}} \left| \psi_i \right\rangle \left\langle  \psi_i\right| \hat{\cal O} \right),
\end{equation}
therefore, our problem reduces to the expectation value in a generic pure state. The trace must be taken over a complete set of independent states (not necessarily quantum states that are normalized to one). Since we are interested in the expectation values of the creation and destruction operators of four-momentum and polarization eigenstates, the most convenient states are $N$-particle and $\bar N$-antiparticle ones. The trace is defined as an integration over the momentum degrees of freedom and a sum over discrete polarizations, namely
\begin{equation}\label{trace_ext}
 \begin{split}
 {\rm tr}\left( \cdots\vphantom{\frac{}{}} \right) &= \sum_r\int\frac{d^3 p}{(2\pi)^3 2 E_{\bf p}} \langle p,r| \cdots |p,r\rangle  +\\
  & \quad +  \sum_{r,s}\int\frac{d^3 p}{(2\pi)^3 2 E_{\bf p}}\frac{d^3 q}{(2\pi)^3 2 E_{\bf q}} \langle p,r,q,s| \cdots |p,r,q,s\rangle + \cdots,
 \end{split}
\end{equation}
and so on, until exausting all the combinations of $N$ particles and $\bar N$ antiparticles. In the last formula the standard definition is used,
\begin{equation}
 |p,r\rangle = \sqrt{2 E_{p}}a^\dagger_r({\bf p})|0\rangle,
\end{equation}
along with the analogous expressions for antiparticles and multiparticle states. The anticommutation relations have the form
\begin{equation}
 \{a_r({\bf p}),a^\dagger_s({\bf p}^\prime)\}=\{b_r({\bf p}),b^\dagger_s({\bf p}^\prime)\} = (2\pi)^3 \delta_{rs}\,\delta^3({\bf p}-{\bf p}^\prime),
\end{equation}
with the normalization
\begin{equation}
 \langle p,r|q,s\rangle = 2E_{\bf p} (2 \pi)^3 \delta_{rs}\,  \delta^3({\bf p}-{\bf p}^\prime).
\end{equation}
It is convenient to introduce the compact notation for multiparticle states 
\begin{equation}
 \begin{split}
 \!\!\!\! |{\underline p},{\underline r};\bar{\underline q},\bar{\underline s}\rangle&=|p_1,r_1,p_2,r_2,\cdots p_N, r_N; {\bar q}_1,{\bar s}_1,{\bar q}_2,{\bar s}_2,\cdots {\bar q}_{\bar N},{\bar s}_{\bar N}\rangle,\\
 \!\!\!\! \int\!\!\![d {\underline p}]^N [d {\underline {\bar q}}]^{\bar N}&= \int\!\!\!\frac{d^3 p_1}{(2\pi)^3 2 E_{{\bf p}_1}}\cdots\frac{d^3 p_N}{(2\pi)^3 2 E_{{\bf p}_N}}\frac{d^3 {\bar q}_1}{(2\pi)^3 2 E_{{\bf {\bar q}}_1}}\cdots\frac{d^3 {\bar q}_{\bar N}}{(2\pi)^3 2 E_{{\bf {\bar q}}_{\bar N}}},
 \end{split}
\end{equation}
where the bar is used to distinguish antiparticle from particle variables. In this way the trace~(\ref{trace_ext}) can be written in a more compact form as
\begin{equation}
 {\rm tr}\left( \vphantom{\frac({}{}}\cdots \right)= \sum_{N,{\bar N}}\sum_{{\underline r},{\underline {\bar s}}} \int\!\!\![d {\underline p}]^N [d {\underline {\bar q}}]^{\bar N} \langle{\underline p},{\underline r};\bar{\underline q},\bar{\underline s}| \cdots |{\underline p},{\underline r};\bar{\underline q},\bar{\underline s}\rangle.
\end{equation}
This compact notation is useful to write the generic quantum state $|\psi\rangle$,
\begin{equation}
 |\psi\rangle = \sum_{N,{\bar N}} \sum_{{\underline r},{\bar {\underline s}}}\int\!\!\![d {\underline p}]^N [d {\underline {\bar q}}]^{\bar N}\; \alpha_{N,{\bar N}}({\underline p},{\underline r};{\underline {\bar q}},{\underline{\bar s}})\,  |{\underline p},{\underline r};\bar{\underline q},\bar{\underline s}\rangle,
\end{equation}
where the complex functions $\alpha_{N,{\bar N}}({\underline p},{\underline r};{\underline {\bar q}},{\underline{\bar s}})$ are partial $N$-particle-$\bar N$-antiparticle wave functions in momentum space. The normalization reads
\begin{equation}\label{norm_psi}
 \begin{split}
  1= \langle\psi|\psi\rangle &= \sum_{N,{\bar N}} \sum_{{\underline r},{\bar {\underline s}}}\int\!\!\![d {\underline p}]^N [d {\underline {\bar q}}]^{\bar N}\; \alpha^*_{N,{\bar N}}({\underline p},{\underline r};{\underline {\bar q}},{\underline{\bar s}}) \alpha_{N,{\bar N}}({\underline p},{\underline r};{\underline {\bar q}},{\underline{\bar s}})=\\
  &= \sum_{N,{\bar N}} \|\alpha_{N,{\bar N}}\|^2,
 \end{split}
\end{equation}
with$ \|\alpha_{N,{\bar N}}\|^2$ being a short-hand notation for the (non-negative\footnote{Being the sum of integrals of a real non-negative weight of the forms $z^* z$.}) sum of integrals
\begin{equation}
  \|\alpha_{N,{\bar N}}\|^2=\sum_{{\underline r},{\bar {\underline s}}}\int\!\!\![d {\underline p}]^N [d {\underline {\bar q}}]^{\bar N}\; \alpha^*_{N,{\bar N}}({\underline p},{\underline r};{\underline {\bar q}},{\underline{\bar s}}) \alpha_{N,{\bar N}}({\underline p},{\underline r};{\underline {\bar q}},{\underline{\bar s}}).
\end{equation}
The tensor product $|\psi\rangle\langle\psi|$, that is,  the projector on the quantum state $|\psi\rangle$ reads
\begin{equation}
\begin{split}
\!\!\!\! |\psi\rangle\langle\psi| &= \sum_{N,{\bar N}} \sum_{{\underline r},{\bar {\underline s}}} \sum_{N^\prime,{\bar N}^\prime} \sum_{{\underline r}^\prime,{\bar {\underline s}}^\prime}\int\!\!\![d {\underline p}]^N [d {\underline {\bar q}}]^{\bar N} [d {\underline p}^\prime]^N [d {\underline {\bar q}}^\prime]^{\bar N}\times \\
 & \qquad  \times \alpha^*_{N^\prime,{\bar N}^\prime}({\underline p}^\prime,{\underline r}^\prime;{\underline {\bar q}}^\prime,{\underline{\bar s}}^\prime)\alpha_{N,{\bar N}}({\underline p},{\underline r};{\underline {\bar q}},{\underline{\bar s}})\,  |{\underline p},{\underline r};\bar{\underline q},\bar{\underline s}\rangle\langle{\underline p}^\prime,{\underline r}^\prime;\bar{\underline q}^\prime,\bar{\underline s}^\prime|.
\end{split}
\end{equation}
Making use of the normalization relations between the states, it is possible to write the trace in a pure state $|\psi\rangle$ of an operator $\hat {\cal O}$ in the compact form
\begin{equation}\label{tr_pure}
 \begin{split}
  &\!\!\!\!{\rm tr}\left( \vphantom{\frac{}{}} \left| \psi \right\rangle \left\langle  \psi\right| \hat{\cal O} \right) =  \sum_{N,{\bar N}} \sum_{{\underline r},{\bar {\underline s}}} \sum_{N^\prime,{\bar N}^\prime} \sum_{{\underline r}^\prime,{\bar {\underline s}}^\prime}\int\!\!\![d {\underline p}]^N [d {\underline {\bar q}}]^{\bar N} [d {\underline p}^\prime]^N [d {\underline {\bar q}}^\prime]^{\bar N} \times \\
  & \times \alpha^*_{N^\prime,{\bar N}^\prime}({\underline p}^\prime,{\underline r}^\prime;{\underline {\bar q}}^\prime,{\underline{\bar s}}^\prime)\alpha_{N,{\bar N}}({\underline p},{\underline r};{\underline {\bar q}},{\underline{\bar s}})\, \langle{\underline p}^\prime,{\underline r}^\prime;\bar{\underline q}^\prime,\bar{\underline s}^\prime|\hat{\cal O} |{\underline p},{\underline r};\bar{\underline q},\bar{\underline s}\rangle.
 \end{split}
\end{equation}
There is a couple of results that can be immediately inferred from the last formula. The first one is that the expectation values of $a^\dagger b^\dagger$ and $b a$, for any momentum and polarization combination, can be non vanishing if and only if the quantum state of the system is in a superposition of states with different particle content. More precisely, only the quantum interference between states that differ exactly by a particle-antiparticle pair can give a non-vanishing contribution ( understanding that the integral over the partial wavefunctions can still simplify and give a vanishing result).

The second observation is that the expectation value of $a^\dagger a$ and $b^\dagger b$ can be simplified. The only combinations that can give a contribution are the ones between states with exactly the same number of particles and the same number of antiparticles. In the following computations, we consider only the term $a^\dagger a$, understanding that the very same transformations hold for antiparticles.

As a particular case of~(\ref{tr_pure}) one can write the expectation value of $a^\dagger_r({\bf p})a_s({\bf p}^\prime)$

\begin{equation}\label{tr_adaggera}
 \begin{split}
  &\!\!\!\!{\rm tr}\left( \vphantom{\frac{}{}} \left| \psi \right\rangle \left\langle  \psi\right| a^\dagger_r({\bf p})a_s({\bf p}^\prime) \right) =  \sum_{N,{\bar N}}\sum_{{\underline t},{\underline t}^{\prime}}\sum_{{\bar {\underline u}},{\bar {\underline u}}^\prime}\int\!\!\![d {\underline k}]^N[d {\underline k}^{\prime}]^N [d {\underline {\bar q}}]^{\bar N}[d {\underline {\bar q}}^\prime]^{\bar N} \times \\
  &\!\!\!\! \times \alpha^*_{N,{\bar N}}({\underline k}^{\prime},{\underline t}^{\prime};{\underline {\bar q}}^\prime,{\underline{\bar u}}^\prime)\alpha_{N,{\bar N}}({\underline k},{\underline t};{\underline {\bar q}},{\underline{\bar u}}) \langle{\underline k}^{\prime},{\underline t}^{\prime};{\underline {\bar q}}^{\prime},{\underline {\bar u}}^{\prime}|a^\dagger_r({\bf p})a_s({\bf p}^\prime) |{\underline k},{\underline t};{\underline {\bar q}},{\underline {\bar u}}\rangle.
 \end{split}
\end{equation}
It is relatively simple to obtain the final formula by making use of the standard anticommutation relations
\begin{equation}
 \begin{split}
  \!\!\!\! &\cdots a^\dagger_r({\bf p})a_s({\bf p}^\prime) \sqrt{2E_{{\bf k}_j}}a^\dagger_{t_j}({\bf k}_j)\cdots = \\
  \!\!\!\!  & \qquad \cdots a^\dagger_r({\bf p}) \sqrt{2E_{{\bf k}_j}} \left(\{ a_s({\bf p}^\prime), a^\dagger_{t_j}({\bf k}_j)\} - a^\dagger_{t_j}({\bf k}_j) a_s({\bf p}^\prime) \right) \cdots = \\
   \!\!\!\! &  \cdots \sqrt{2E_{{\bf k}_j}}\left( a^\dagger_{t_j}({\bf k}_j)  a^\dagger_r({\bf p})a_s({\bf p}^\prime) + a^\dagger_r({\bf p}) (2\pi)^3\delta_{s t_j}\delta^3({\bf k}_j -{\bf p}^\prime)  \right) \cdots = \\
   \!\!\!\! \!\!\! & \!\!\!\!\! \cdots\left[ \sqrt{2E_{{\bf k}_j}} a^\dagger_{t_j}({\bf k}_j) \left(  a^\dagger_r({\bf p})a_s({\bf p}^\prime)  \right) + \sqrt{2E_{{\bf p}^\prime}} (2\pi)^3\delta_{s t_j}\delta^3({\bf k}_j -{\bf p}^\prime) a^\dagger_r({\bf p})\right]\cdots =\\ \\
 \!\!\!\!\!\!\! & \!\!\!\!\!\!\!\! =\cdots\left[ \sqrt{2E_{{\bf k}_j}} a^\dagger_{t_j}({\bf k}_j) \left(  a^\dagger_r({\bf p})a_s({\bf p}^\prime)  \right) + \right.\\
  & \qquad \left. + \sqrt{\frac{2E_{{\bf p}^\prime}}{2E_{{\bf p}}}} (2\pi)^3\delta_{s t_j}\delta^3({\bf k}_j -{\bf p}^\prime)  \sqrt{2E_{{\bf p}}}a^\dagger_r({\bf p})\right]\cdots 
 \end{split}
\end{equation}
In other words, even if $ a^\dagger_r({\bf p})a_s({\bf p}^\prime) $ doesn't commute with the creation operators, it is possible to "move it to the right". However, each time we do that we have to add a new state, with a delta between the $j$'th degrees of freedom and the destruction operator $a_s({\bf p}^\prime)$, a numerical factor $(2\pi)^3\sqrt{E_{{\bf p}^\prime}/E_{\bf p}}$ and a substitution of the momentum and polarization at the $j$'th place with the ones related to the creation operator $ a^\dagger_r({\bf p})$. After moving to the right all the particle creation operators, $ a^\dagger_r({\bf p})a_s({\bf p}^\prime) $ commutes with the creation operators of the antiparticles (if present). 

In the end, after making use of the normalization of the eigenstates we find
\begin{equation}\label{tr_adaggera_final}
 \begin{split}
  &\!\!\!\!{\rm tr}\left( \vphantom{\frac{}{}} \left| \psi \right\rangle \left\langle  \psi\right| a^\dagger_r({\bf p})a_s({\bf p}^\prime) \right) =  0 +\\
  &\quad + \frac{1}{\sqrt{2E_{\bf p}2E_{{\bf p}^\prime}}}  \sum_{{\bar N},N>0}\sum_{j+1}^N\sum_{{\underline t}-t_j}\sum_{{\bar {\underline u}}}\int\!\!\![d {\underline k}]^{(N-j)} [d {\underline {\bar q}}]^{\bar N}\times \\
  &\!\!\!\! \times \alpha^*_{N,{\bar N}}({\underline k}-{\bf k}_j,{\bf p}, {\underline t}-t_j, r;{\underline {\bar q}},{\underline{\bar u}})\alpha_{N,{\bar N}}({\underline k}-{\bf k}_j,{\bf p}^\prime, {\underline t}-t_j, s;{\underline {\bar q}},{\underline{\bar u}}).
 \end{split}
\end{equation}
The notation $\sum_{{\underline t}-t_j}\int d {[\underline k}]^{(N-j)}$ means that the integral and the sum is over all the particle degrees of freedom except for the $j$'th. In the similar way $\alpha_{N,{\bar N}}({\underline k}-{\bf k}_j,{\bf p}, {\underline t}-t_j, r;{\underline {\bar q}},{\underline{\bar u}})$ is a shorthand notation for the (partial) wavefunction with the $j$'th degrees of freedom fixed to the momentum ${\bf p}$ and polarization $r$. 

The formula~(\ref{tr_adaggera_final}) has many interesting consequences. Besides the expected vanishing expectation value for purely antiparticle states, one can immediately check that the expectation value of $a^\dagger_r({\bf p})a_r({\bf p}) $ is non-negative, since it is a series of integrals and sums of squares. Moreover, as one could expect, it is linked to the average number of particles. Indeed, the expression
\begin{equation}
 \sum_r \int \frac{d^3 p}{(2 \pi)^3} {\rm tr}\left( |\psi\rangle\langle \psi | a^\dagger_r({\bf p})a_r({\bf p}) \right) = \sum_N N \sum_{\bar N}\| \alpha_{N,{\bar N}}\|^2,
\end{equation}
exactly gives the average number of particles in the state $|\psi\rangle$
because of the normalization~(\ref{norm_psi}). More interestingly, the expectation value of $a^\dagger_r({\bf p})a_s({\bf p})$ (same momentum, different polarization) performs the role of a momentum dependent spin density matrix. The momentum integral of the trace is proportional to the average number of particles, but the matrix itself is sensitive to polarization in the $r , s$ indices and can be used to obtain the average number of particles, per momentum cell, for some polarization states.

All these arguments do not change if one reinserts the classical probabilities ${\sf P}_i$ from~(\ref{density}) and deals with mixed states. The classical fluctuations do not change the properties of the spin density matrix, like the non-negative diagonal elements and normalization of the trace (after dividing by $(2\pi)^3$ and integrating over momentum, like for the pure states) does not change the average number of particles.



\begin{thebibliography}{99.}

\bibitem{Busza:2018rrf}
  W.~Busza, K.~Rajagopal and W.~van der Schee,
  Ann.\ Rev.\ Nucl.\ Part.\ Sci.\  {\bf 68} (2018) 339.

\bibitem{Romatschke:2017ejr}
P.~Romatschke and U.~Romatschke,
doi:10.1017/9781108651998, arXiv:1712.05815 [nucl-th].
  
\bibitem{Florkowski:2017olj}
  W.~Florkowski, M.~P.~Heller and M.~Spalinski,
  Rept.\ Prog.\ Phys.\  {\bf 81} (2018) 046001.

\bibitem{Borsanyi:2010bp}
  S.~Borsanyi {\it et al.} [Wuppertal-Budapest Collaboration],
  JHEP {\bf 1009} (2010) 073.
  
\bibitem{Bazavov:2018mes}
  A.~Bazavov {\it et al.} [HotQCD Collaboration],
  Phys.\ Lett.\ B {\bf 795} (2019) 15.
 
\bibitem{Miller:2007ri}
  M.~L.~Miller, K.~Reygers, S.~J.~Sanders and P.~Steinberg,
  Ann.\ Rev.\ Nucl.\ Part.\ Sci.\  {\bf 57} (2007) 205.
  
\bibitem{Gelis:2010nm}
  F.~Gelis, E.~Iancu, J.~Jalilian-Marian and R.~Venugopalan,
  Ann.\ Rev.\ Nucl.\ Part.\ Sci.\  {\bf 60} (2010) 463.  
  
\bibitem{Cooper:1974mv}
F.~Cooper and G.~Frye, Phys.\ Rev.\ D {\bf 10} (1974) 186.
  
\bibitem{de_Groot}
S. R. de Groot, W. A. van Leeuwen, and Ch. G. van Weert, {\it Relativistic Kinetic Theory}, North–
Holland, Amsterdam (1980).

\bibitem{Wang:2017jpl}
  Q.~Wang,
  Nucl.\ Phys.\ A {\bf 967} (2017) 225

\bibitem{Huang:2020xyr}
  X.~G.~Huang,
  arXiv:2002.07549 [nucl-th].

\bibitem{Becattini:2020ngo}
  F.~Becattini and M.~A.~Lisa,
  arXiv:2003.03640 [nucl-ex].

\bibitem{Becattini:2020sww}
  F.~Becattini,
  arXiv:2004.04050 [hep-th].
  
\bibitem{Florkowski:2010zz}
  W.~Florkowski, {\it Phenomenology of Ultra-Relativistic Heavy-Ion Collisions}, World Scientific, 
  Singapore (2010).
  
\bibitem{Itzykson:1980rh} 
  C.~Itzykson and J.~B.~Zuber,
  {\it Quantum Field Theory},
   McGraw-Hill, New York (1980). 

\bibitem{Florkowski:2018ahw}
  W.~Florkowski, A.~Kumar and R.~Ryblewski,
  Phys.\ Rev.\ C {\bf 98} (2018) 044906.
  
\bibitem{Weickgenannt:2019dks}
  N.~Weickgenannt, X.~L.~Sheng, E.~Speranza, Q.~Wang and D.~H.~Rischke,
  Phys.\ Rev.\ D {\bf 100} (2019) 56018.
  
\bibitem{Liu:2020flb}
  Y.~C.~Liu, K.~Mameda and X.~G.~Huang,
  arXiv:2002.03753 [hep-ph].

\bibitem{Hehl:1976vr}
  F.~W.~Hehl,
  Rept.\ Math.\ Phys.\  {\bf 9} (1976) 55.
  
\bibitem{Florkowski:2018fap}
  W.~Florkowski, R.~Ryblewski and A.~Kumar,
  Prog.\ Part.\ Nucl.\ Phys.\  {\bf 108} (2019) 103709.
  
  \bibitem{Weickgenannt:2020aaf}
N.~Weickgenannt, E.~Speranza, X.~l.~Sheng, Q.~Wang and D.~H.~Rischke,
arXiv:2005.01506 [hep-ph].
  
\bibitem{Speranza:2020ilk}
    E.~Speranza, and N.~Weickgenannt,
    arXiv:2007.00138 [nucl-th].

  
\bibitem{Becattini:2013fla}
  F.~Becattini, V.~Chandra, L.~Del Zanna and E.~Grossi,
  Annals Phys.  {\bf 338} (2013) 32.
 
\bibitem{Becattini:2018duy}
  F.~Becattini, W.~Florkowski and E.~Speranza,
  Phys.\ Lett.\ B {\bf 789} (2019) 419
  
\bibitem{Mathisson}
M.~Mathisson, Acta Phys. Pol. {\bf 6} (1937) 163.

\bibitem{Weyssenhoff}
J.~Weyssenhoff and A. Raabe, Acta Phys. Pol. {\bf 9} (1947) 7.

\end{thebibliography}
\end{document}